\documentclass[11pt,a4paper]{article}
\usepackage{amsmath,amssymb,amsfonts,slashed}
\usepackage[text={176mm,240mm},centering]{geometry}
\usepackage{float}
\usepackage{graphicx}
\usepackage{multirow}
\usepackage{color}
\usepackage{cite}
\usepackage[colorlinks=true,linkcolor=red,breaklinks=true,urlcolor=blue,citecolor=blue]{hyperref}

\newcommand \bra [1] {\langle {#1}\vert}
\newcommand \ket [1] {\vert {#1}\rangle}
\newcommand \inner [2] {\langle {#1}\vert {#2}\rangle}
\newcommand{\tbreak}[2]{\begin{tabular}{@{}#1@{}}#2\end{tabular}}

\def\be{\begin{equation}}
\def\ee{\end{equation}}
\def\ba{\begin{eqnarray}}
\def\ea{\end{eqnarray}}
\def\3P0{{}^3P_0}

\begin{document}
\title{Coupled-Channel Effects for the Bottomonium with Realistic Wave Functions}

\date{\today}

\author{
     Yu Lu$^{1,3}$
                         \footnote{Email address:\texttt{luyu211@ihep.ac.cn} }~ ,
     Muhammad Naeem Anwar $^{2,3}$
                        \footnote{Email address:\texttt{naeem@itp.ac.cn} }~ ,
      Bing-Song Zou$^{2,3}$
                        \footnote{Email address:\texttt{zoubs@itp.ac.cn}}
       \\[2mm]
      {\it\small$^1$Institute of High Energy Physics, Chinese Academy of Sciences, Beijing 100049, China,}\\
      {\it\small$^2$CAS Key Laboratory of Theoretical Physics, Institute of Theoretical Physics,}\\
      {\it\small  Chinese Academy of Sciences, Beijing 100190, China}\\
      {\it\small$^3$University of Chinese Academy of Sciences, Beijing 100049, China}\\
}

\maketitle

\begin{abstract}
With Gaussian expansion method (GEM),
realistic wave functions are used to calculate coupled-channel effects for the bottomonium under the framework of $\3P0$ model.
The simplicity and accuracy of GEM are explained.
We calculate the mass shifts, probabilities of the $B$ meson continuum, $S-D$ mixing angles,
strong and dielectric decay widths.
Our calculation shows that both $S-D$ mixing and the $B$ meson continuum can contribute to
the suppression of the vector meson's dielectric decay width.
We suggest more precise measurements on the radiative decays of $\Upsilon(10580)$ and $\Upsilon(11020)$ to distinguish these two effects.
The above quantities are also calculated with simple harmonic oscillator (SHO) wave function approximation for comparison.
The deviation between GEM and SHO indicates that it is essential to treat the wave functions accurately for near threshold states.
\end{abstract}


\thispagestyle{empty}

\section{Introduction}
Heavy quarkonium is a multiscale system covering all regimes of quantum chromodynamics (QCD) which make it an ideal place to study strong interactions~\cite{Brambilla:2010cs}.
Despite the success of QCD in high energy region, due to asymptotic freedom,
nonperturbative effect dominates at low energies and brings problems to perturbative calculation.
One tool to study this nonperturbative effect is lattice QCD.
However, due to its huge calculation work, it is still unable to calculate all the physical quantities with the current computation power.
Another important approach is to develop various phenomenological models.
Among these phenomenological models, the quark model is a prominent one.
Under the quark model framework,
various types of interactions have been suggested by various groups,
and they have achieved many impressive successes (see e.g. Refs.~\cite{Martin:1980rm,Bertlmann:1979zs,Eichten:1978tg,Buchmuller:1980su,Godfrey:1985xj}).
However, these potential models cannot be the whole story.
One important missing ingredient is the mechanism to generate quark-antiquark pairs
which enlarge the Fock space of the initial state,
i.e. the initial state contains multiquark components.

These multiquark components will change the Hamiltonian of the potential model,
causing mass shift and mixing between states with the same quantum numbers
or directly contributing to open channel strong decay if the initial state is above threshold.
These consequences can be summarized as unquenched effects or coupled-channel effects.
Coupled-channel effects have been considered at least 30 years ago by
T\"{o}rnqvist \emph{et al.} in Refs.~\cite{Heikkila:1983wd,Ono:1983rd,Tornqvist:1984fy,Ono:1985eu,Ono:1985jt};
they extended the quark model to be unquenched quark model.

Despite the fact that the underlying quark pair creation mechanism is not fully understood up to now,
still there are different phenomenological models to decode the mystery,
such as $\3P0$ model~\cite{Micu:1968mk,LeYaouanc:1972ae,LeYaouanc:1973xz},
flux-tube breaking model~\cite{Kokoski1987,Close:1994hc},
microscopic decay models~\cite{Eichten:1978tg,Eichten:1979ms,Ackleh:1996yt}.
Among these, the most simple and successful one is the $\3P0$ model,
where the generated light quark pair share the same quantum number as vacuum.

Even though $\3P0$ model is extensively studied by many people, 
almost all the calculations are using SHO wave function approximation to simplify the calculation
(see e.g.~\cite{Barnes:2002mu,Ackleh:1996yt,Barnes:1996ff,Yang:2009fj,Liu:2011yp,Ferretti:2013vua,Chen:2007xf,An:2012kj,An:2013zoa}).
A simple yet powerful method to handel the wave function precisely is still not widely known.
We propose using the Gaussian expansion method (GEM)~\cite{Hiyama:2003cu} to accurately evaluate the wave function convolution.

There have been already some works related to GEM.
In Refs.~\cite{Segovia:2008zz,Segovia:2011dg,Segovia:2012cd,Segovia:2013wma,Segovia:2013kg,
Yang:2010sf,Deng:2010zzd,Yang:2011rp,Deng:2012nn,Gao:2012zza,Deng:2012wi,Deng:2013aca,Segovia:2015dia,Segovia:2016xqb},
GEM is adopted to calculate the wave functions under variational method approach.
Coupled-channel effects with GEM are only studied for some specific cases,
such as $X(3872)$ and $P$ wave $D_s$ mesons~\cite{Ortega:2009hj,Ortega:2016mms}.
In Refs.~\cite{Wang:2014lml,Wang:2014voa,Chen:2014ztr},
the authors also use GEM to calculate the spectrum and open channel strong decays of light mesons and some specific charmonia,
where the coupled channel induced mass shift is not considered.
Even though the mass shifts can be partly absorbed by redefining the potential,
the potential model cannot describe near-threshold effects~\cite{Li:2009ad}.
We want to emphasize that the mass shifts and open channel strong decays are directly correlated by coupled-channel effects,
so it is essential to evaluate them under the same framework and calculate them precisely.

So far, a precise evaluation and a thorough discussion of the coupled-channel effects are still missing,
and the validity of the SHO approximation is yet to be clarified.
In this paper, we fill these gaps by a thorough discussion of coupled-channel effects for the bottomonium
and we also predict some important results on the dielectric and radiative decays of vector mesons
which are going to be tested by experiments.

The paper is organized as follows.
In Sec.~\ref{modelSec}, we explain the details of Cornell potential and $\3P0$ model,
where we deduce the formula of mass shift, open channel strong decay width and $S-D$ mixing.
In  Sec.~\ref{calSec}, we focus on the calculation details and GEM,
where the advantages of GEM are elucidated and the procedure to fit the wave function is explained.
Sec.~\ref{resultSec} is devoted to discussing the possible impacts of coupled-channel effects for the bottomonium
on the spectrum, open channel strong decays, probabilities of the $B$ meson continuum,
the $S-D$ mixing and the vector meson's dielectric and radiative decays.
We also explicitly show the deviation between GEM and SHO approximation.
Finally, we give a short summary of this work in Sec.~\ref{summary}.

\section{Theoretical Framework}\label{modelSec}
\subsection{Cornell Potential Model}
As the quenched limit,
the wave functions for the heavy quarkonium are obtained by solving the Schr\"{o}dinger equation
with the well-known Cornell potential~\cite{Eichten:1978tg,Eichten:1979ms}
\be
V(r)=-\frac{4}{3} \frac{\alpha}{r}+\lambda r+c,
\ee
where $\alpha, \lambda$ and $c$ stand for the strength of color Coulomb potential,
the strength of linear confinement and mass renormalization, respectively.
To restore the hyperfine or fine structures of the bottomonium,
we use the following form of the spin dependent interactions

\be
V_{s}(r)=\left(\frac{2\alpha}{m^2_b r^3}-\frac{\lambda}{2m^2_b r}\right)\vec{L}\cdot \vec{S}
+\frac{32\pi \alpha}{9m_b^2}\tilde{\delta}(r) \vec{S}_b\cdot \vec{S}_{\bar{b}}
+\frac{4\alpha}{m^2_b r^3}\left(\frac{\vec{S}_b\cdot \vec{S}_{\bar{b}}}{3}
+\frac{(\vec{S}_b\cdot\vec{r}) (\vec{S}_{\bar{b}}\cdot \vec{r})}{r^2}\right),
\label{finestructure}
\ee
where $\vec{L}$ denotes relative orbital angular momentum,
$\vec{S}=\vec{S}_b+\vec{S}_{\bar{b}}$  is the total spin of the $b$ quark pairs
and $m_b$ is the $b$ quark mass.
Since the nonrelativistic expansion will fail if two composite quarks are very close to each other,
instead of the Dirac $\delta$ function in the second term,
we use the smeared delta function,
which can be written as
$\tilde{\delta}(r)=(\sigma/\sqrt{\pi})^3 e^{-\sigma^2 r^2}$~\cite{Barnes:2005pb,Li:2009ad}.
The Hamiltonian of the Schr\"{o}dinger equation in quenched limit is represented as
\be
H_0 =2m_b+\frac{p^2}{m_b}+V(r)+V_{s}(r).
\ee
We treat the spin dependent term as a perturbation and
the spatial wave functions are obtained by solving Schr\"{o}dinger  equation numerically using Numerov's method~\cite{Numerov:1927}.

\subsection{$\3P0$ Model and Coupled-Channel Effects}
For the coupled channel calculation,
we adopt the widely used ${}^3P_0$ model or quark pair creation model,
which is first proposed by L. Micu~\cite{Micu:1968mk} in 1969 and then extended by A. Le Yaouanc \emph{et al.} in 1970s~\cite{LeYaouanc:1972ae,LeYaouanc:1973xz}.
In this model, the generated quark pairs have vacuum quantum number $J^{PC}=0^{++}$.
After simple arithmetic, one can conclude that the relative orbital angular momentum and total spin are both equal to 1.
In the notation of ${}^{2S+1}L_J$, one should write it as $\3P0$ which explains the model's name.

The interaction Hamiltonian can be expressed as
\be
H_I=2 m_q \gamma \int d^3x \bar{\psi}_q \psi_q,
\ee
where $m_q$ is the produced quark mass,
and $\gamma$ is the dimensionless coupling constant.
Since the probability to generate heavier quarks is suppressed,
we use the effective strength $\gamma_s=\frac{m_q}{m_s}\gamma$ in the following calculation,
where $m_q=m_u=m_d$ is the constituent quark mass of up (or down) quark and $m_s$  is strange quark mass.

\begin{figure}[h]
  \centering
  \includegraphics[width=0.6\textwidth]{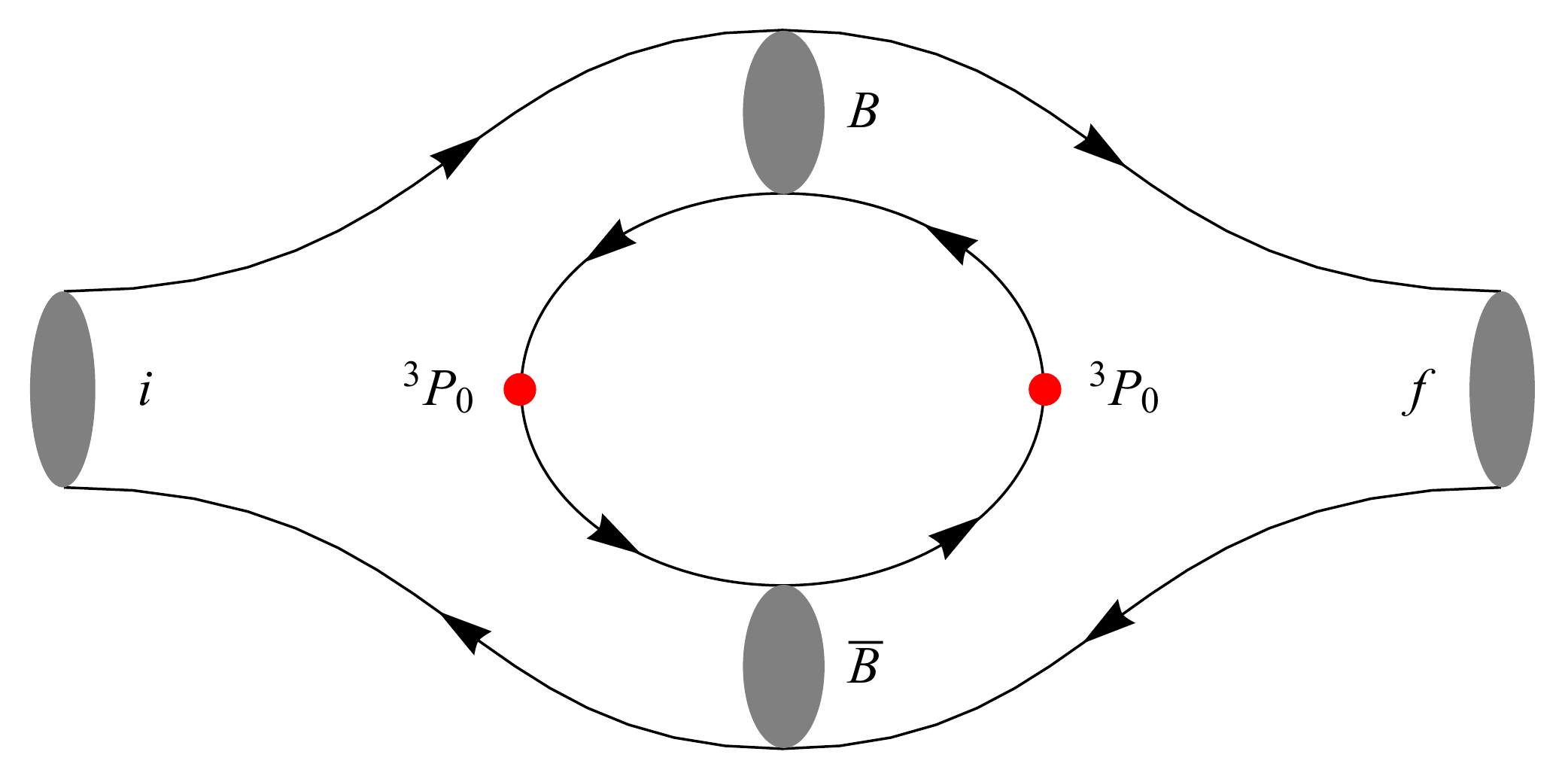}\\
  \caption{Sketch of coupled-channel effects in $\3P0$ model.
   $i$ and $f$ respectively denote the initial and final states with same $J^{PC}$ and
   $B\bar{B}$ stands for all possible $B$ meson pairs.
  }
 \label{ccDiagram}
\end{figure}

The $\3P0$ Hamiltonian induces not only open-flavor strong decays of the heavy quarkonium above threshold,
but also coupled-channel effects.
As sketched by Fig.~\ref{ccDiagram}, the experimentally observed state should be a mixture of pure quarkonium state (bare state) and $B$ meson continuum.
Put it in to formula, the physical or experimentally observed state $\ket{A} $ should be expressed as
\be
\ket {A}=c_0 \ket{\psi_0} +\sum_{BC} \int d^3p\, c_{BC}(p) \ket{BC;p},
\ee
where $c_0$ and $c_{BC}$ stand for the normalization constants of the bare state and $B$ meson continuum, respectively.
$\ket {\psi_0}$ is normalized to 1 and $\ket {A}$ is also normalized to 1 if it lies below $B\bar{B}$ threshold.
$\ket{BC;p}$ is normalized as $\inner{BC;p_1}{B'C';p_2}=\delta^3(p_1-p_2)\delta_{BB'}\delta_{CC'}$,
where $p$ is the momentum of $B$ meson in $\ket{A}$'s rest frame.
Combining the Cornell potential and the dynamics of quark pair generation,
we get the full Hamiltonian,
\be
  H=H_0+H_{BC}+H_I,
\ee
with the following relations
\ba
  H_0 \ket{\psi_0} &=& M_0 \ket {\psi_0} \\
  H_0 \ket{BC;p} &=& 0 \\
  H_{BC} \ket{\psi_0} &=& 0\\
  H_{BC} \ket{BC;p} &=& E_{BC}\ket{BC;p}\\
  H \ket{A}&=& M \ket{A},\label{Eigen}
\ea
where $M_0$ is the bare mass of the bottomonium and can be solved directly from Schr\"{o}dinger equation.
The interaction between $B$ mesons is neglected,
so the energy of meson continuum can be expressed as
$E_{BC}=\sqrt{m_B^2+p^2}+\sqrt{m_C^2+p^2}.$

When Eq.~(\ref{Eigen}) is projected onto each component, we immediately get
\ba
\bra{\psi_0} H \ket{\psi}=c_0 M=c_0 M_0+ \int d^3p \, c_{BC}(p) \bra {\psi_0} H_I \ket{BC;p}, \label{eqnC2}\\
\bra{BC;p} H \ket{\psi}=c_{BC}(p)M=c_{BC}(p) E_{BC}+c_0 \bra{BC;p} H_I \ket{\psi_0}. \label{eqnC4}
\ea
Solve $c_{BC}$ from Eq.~(\ref{eqnC4}), 
substitute back to Eq.~(\ref{eqnC2}) and eliminate the $c_0$ on both sides,
we get a integral equation
\be
M=M_0+\Delta M, \label{intEqn}
\ee
where
\be
\Delta M=\sum_{BC}\int d^3p\, \frac{\vert \bra {BC;p} H_I \ket{\psi_0} \vert ^2}{M-E_{BC}-i\epsilon}. \label{mShift}
\ee
The sum of $BC$ is restricted to the ground state $B_{(s)}$ mesons, i.e. $B\bar{B}, B\bar{B}^{*}+h.c., B^{*}\bar{B}^{*}, B_s\bar{B}_s, B_s\bar{B}^{*}_s+h.c., B^{*}_s\bar{B}^{*}_s$.
Note that the $i\epsilon$ term is added to handle the situation when $m_A> m_B+m_C$.
In this case, $\Delta M$ will pick up an imaginary part
\be
\mathrm{Im} (\Delta M)=\sum_{BC} \pi P_B \frac{E_B E_C}{m_A} \vert \bra {BC;P_B} H_I \ket{\psi_0} \vert ^2,\label{decay}
\ee
which is equal to one half of the the decay width.
$P_B$ and $ E_B$ respectively denote the momentum and energy of $B$ meson.
The wave function overlap integration lies in the term
\be
\bra{BC;P_B} H_I \ket{\psi_0}=
\sum_{\text{polarization}}\int d^3k
\phi_0(\vec{k}+\vec{P}_B) \phi_B^*(\vec{k}+x_B \vec{P}_B)\phi_C^*(\vec{k}+x_C \vec{P}_B)
|\vec{k}| Y_1^m(\theta_{\vec{k}},\phi_{\vec{k}}),\label{overlap}
\ee
where $x_B=m_4/(m_1+m_4), x_C=m_3/(m_2+m_3)$ and
$m_1=m_2=m_Q, m_3=m_4$ respectively denote the $b$ quark and the light quark mass.

Once $M$ is solved, the coefficient of different components can be worked out either.
For states below threshold, the normalization condition $\ket{A}$ can be rewritten as
\be
|c_0|^2+\int d^3p|c_{BC}|^2=1
\ee
after the substitution of $c_{BC}$, we get the probability of the $b\bar{b}$ component
\be
P_{b\bar{b}}:=\vert c_0 \vert ^2=1/\left(1+\sum_{BC LS}\int_0^\infty dp \frac{p^2\vert \mathcal{M}^{LS}\vert^2}{(M-E_{BC})^2}\right), \label{1prob}
\ee
where $\vert \mathcal{M}^{LS}\vert^2$ is represented as
\be
\vert \mathcal{M}^{LS}\vert^2=\int d\Omega_B \, \vert \bra {BC;P_B} H_I \ket{\psi_0} \vert ^2.
\ee

\subsection{Coupled Channel Induced $S-D$ Mixing}
From the quark model's perspective,
the spatial wave functions of $J^{PC}=1^{--}$ family can be both $S$ and $D$ wave.
It is natural to expect that the experimentally observed vector states are the mixing of $S$ and $D$ waves.
As in the case of conventional meson coupling with $B\bar{B}$ continuum,
we rewrite it into a matrix form
\be
\left(
  \begin{array}{cc}
    M_0 & \int d^3p \bra{\psi_0} H_I \ket{BC} \\
    \bra{BC} H_I \ket{\psi_0} & E_{BC} \\
  \end{array}
  \right)
  \left(
  \begin{array}{c}
   c_0\\
    c_{BC} \\
  \end{array}
  \right)
=
M
\left(
  \begin{array}{c}
   c_0\\
    c_{BC} \\
  \end{array}
  \right),
\ee
where the integration part should be understood as a formal notation,
and one needs to insert all the $p$ dependent part into the integral.
For example, in the above case,
one may naively get the following form after diagonalization
\be
(M-M_0)(M-E_{BC})=\int d^3p |\bra{\psi_0}H_I\ket{BC}|^2.
\ee
However, the correct form should be understood as Eq.(\ref{mShift}),
where the $(M-E_{BC})$ term is in the integration.

The advantage of the matrix form is that one can easily see its structure and it can be easily generalized to $S-D$ mixing case.
Under the assumption that $(n+1)S$ mix only with $nD$, we have
\be
\left(
  \begin{array}{ccc}
    M_S^0 &H_T & \int d^3p \bra{\psi_S} H_I \ket{BC}  \\
    H_T &M_D^0 & \int d^3p \bra{\psi_D} H_I \ket{BC}  \\
    \bra{BC} H_I \ket{\psi_S} & \bra{BC} H_I \ket{\psi_D} & E_{BC} \\
  \end{array}
  \right)
  \left(
  \begin{array}{c}
   c_S\\
   c_D\\
   c_{BC} \\
  \end{array}
  \right)
=
M
  \left(
  \begin{array}{c}
   c_S\\
   c_D\\
   c_{BC} \\
  \end{array}
  \right).
\ee

The $S-D$ mixing induced by tensor part of the potential is so small,
typically around $0.8^\circ$ in our calculation (see also the Appendix A in Ref.~\cite{Badalian:2009bu}),
so its quite reasonable to set $H_T=0$.
After this approximation, one can reexpress $c_{BC}$ in terms of $c_S,c_D$ and easily get
\be
\left(
  \begin{array}{cc}
    M_S^0+\Delta M_S & \Delta M_{SD} \\
    \Delta M_{DS}  & M_D^0+\Delta M_D\\
  \end{array}
  \right)
  \left(
  \begin{array}{c}
   c_S\\
   c_D\\
  \end{array}
  \right)
=
M
  \left(
  \begin{array}{c}
   c_S\\
   c_D\\
  \end{array}
  \right),
  \label{SDMixing}
\ee
where
\ba
\Delta M_f&=&\int d^3p \frac{|\bra{\psi_f}H_I\ket{BC}|^2}{M-E_{BC}-i\epsilon} \qquad (f=S,D),\\
\Delta M_{SD}=\Delta M_{DS}^*&=&\int d^3p \frac{\bra{\psi_S}H_I\ket{BC}\bra{BC}H_I\ket{\psi_D}}{M-E_{BC}-i\epsilon}.
\ea
From the above equation, both the mass and the relative ratio $c_S/c_D$ can be worked out.
For states below threshold, the probability can be solved once the mass is known,
which is a generalization of Eq.~(\ref{1prob})
\be
|c_S|^2+|c_D|^2+\sum_{BC} \int d^3p \frac{1}{(M-E_{BC})^2}(|c_S|^2 H_{S,BC}^2+|c_D|^2 H_{D,BC}^2+2 \text{Re}[c_S c_D^* H_{S,BC}H_{BC,D}])=1,
\label{SDMixingRatio}
\ee
where $H_{f,i}$ stands for$\bra{f}H_I\ket{i}$.

One will get a complex solution of $M=M_{\text{BW}}+i \Gamma/2$ if $M_{\text{BW}}>m_B+m_{\bar{B}}$,
where $M_{\text{BW}}$ represents the Breit-Wigner mass of the resonance,
and $\Gamma$ is the decay width after considering $S-D$ mixing.
As a cross check, one can also calculate the decay width directly with the following formula,
\be
\Gamma_{SD}=2\Big(|c_S|^2 \mathrm{Im}(\Delta M_{S})+|c_D|^2 \mathrm{Im}(\Delta M_{D})+2\mathrm{Re}\big(c_S^* c_D \mathrm{Im}(\Delta M_{SD})\big)\Big).
\ee

Eq.~(\ref{SDMixing}) is much more difficult to solve than the Eq.~(\ref{mShift}).
The method we use to solve this equation will be discussed in the next section.

\section{Parameter Selection and Gaussian Expansion Method}\label{calSec}
\subsection{Parameter Selection}\label{paraTune}
As a first step, we tune the wave functions to be consistent with the dielectric decay widths of $\Upsilon(nS)$ for $n\le3$.
The parameters are given in Table~\ref{paraTab}.
Theoretically, dielectric decay widths can be expressed as~\cite{VanRoyen:1967nq,Moxhay:1983vu,Barbieri:1979be,Rosner:2001nm}
\be
\Gamma_{ee}=\beta\frac{4\alpha^2 e^2_b}{M^2_{nS}}|c_S \, R_{nS}(0)+c_D \frac{5}{2\sqrt{2}m_b^2}R_{nD}''(0)|^2 \\
\label{eeFormula}
\ee
where $\beta=(1-16\alpha_s/3\pi)$ is the QCD radiative correction,
and $e_b=-1/3$ is the $b$ quark charge in the unit of electron charge.
$R_{nS}(0) $ denotes the radical $S$ wave function at the origin,
and $R_{nD}''(0)$ is the second derivative of the radical $D$ wave function at the origin.
$c_S$ and $ c_D$ respectively denote the normalization coefficients before $S$ and $D$ wave.
Note that from the perspective of coupled channels,
they are not restricted to be real-valued and $|c_S|^2+|c_D|^2\ne 1$.
For below threshold states, the correct normalization is given by Eq.~(\ref{SDMixingRatio}).
Nevertheless, if the imaginary part of the $\Delta M$ are neglected in Eq.~(\ref{SDMixing}),
the corresponding solutions will be real,
and one can easily get the feel of how big the mixing is by defining
$\tan\theta:=|c_S/c_D|$ for $S$ wave dominate states and $\tan\theta:=|c_D/c_S|$ for $D$ wave dominate states.

There is also an argument for the above formula that
the QCD corrections of higher order may be important~\cite{Badalian:2009bu},
thus $\beta$ has to be treat as an effective constant.
So in order to reduce the parameter's uncertainty,
we tune the wave functions to reproduce $\Gamma_{ee}(nS)/\Gamma_{ee}(1S), (n=2,3)$(see Fig.~\ref{eeDecayPlot}).

\begin{table}[H]
  \renewcommand\arraystretch{1.1}
  \centering
\begin{tabular}{cccccccccc}
 \hline\hline
$\alpha=0.34$   & $\lambda=0.22\text{GeV}^2$    &$c=0.435\text{GeV} $\\
$m_b=4.5\text{GeV}$ & $m_u=m_d=0.33\text{GeV}$& $m_s=0.5\text{GeV}$\\
$\sigma=3.838\text{GeV}$& $\gamma=0.205$ \\
 \hline \hline
\end{tabular}
\caption{The parameters used in our calculation.
These parameters are chosen to reproduce the dielectric decay widths of $\Upsilon(nS), n=1,2,3$,
which are shown in Fig.~\ref{eeDecayPlot}.
Due to the implicit treatment of color and flavor degrees of freedom,
these factors do not show up in our calculations.}
\label{paraTab}
\end{table}

\subsection{Gaussian Expansion Method}

There are at least two ways to solve Eq.~(\ref{mShift}).
The first one is recursion method,
which is based on the observation that the mass shift is expected to be small compared with the bare mass.
i.e. set $M=m_0$ as the first step and do the integration in Eq.~(\ref{mShift}) to get the mass shift,
then set $M=m_0+\Delta m$ again and so on until the result converges.

One can even make a further approximation and only do the first step recursion.
However, this method only applies to the single channel mass shift formula~(\ref{mShift}).
In $S-D$ mixing cases, such as Eq.~(\ref{SDMixing}),
the mass difference between $M_S^0+\Delta M_S, M_D^0+\Delta M_D$ is small,
even a small error in the off-diagonal term in Eq.~(\ref{SDMixing}) will ruin the prediction of the $S-D$ mixing angle.

The second way is to solve the equation by brute-force,
i.e. for the energy ranges we are interested in,
work out all of the integrations in Eq.~(\ref{mShift}) or Eq.~(\ref{SDMixing}) at specific energy point.
We use this method despite of its huge calculation work.
The benefit is that we can extract a lot of information about the wave function's impact on the mass shift.
One can also change the $\3P0$'s coupling constant $\gamma$ or the mass renormalization constant $c$ to see the possible consequences.

The high precision work will not be convincing if there is no way to precisely evaluate the integration,
which has a key ingredient --- the wave function.
One can indeed evaluate the amplitude pure numerically as the authors do in Refs.~\cite{Ono:1981pt,Li:2009ad},
however, we still want analytic expressions which is more convenient
if we want to change the parameters and then repeat the calculations.

In order to achieve that,
various groups approximate the wave functions by simple harmonic oscillators (SHOs) approximation (see e.g.~\cite{Barnes:2002mu,Ackleh:1996yt,Barnes:1996ff,Ferretti:2013vua,Liu:2011yp}).
The oscillator parameters $\beta$s are usually settled down by requiring that
the root mean square radii to be equal to the initial states~\cite{Kalashnikova:2005ui,Godfrey:2015dia,Zhou:2013ada}
or maximizing its overlap with the numerical wave function~\cite{Ackleh:1996yt}.

\begin{figure}[h]
  \centering
  \includegraphics[width=0.8\textwidth]{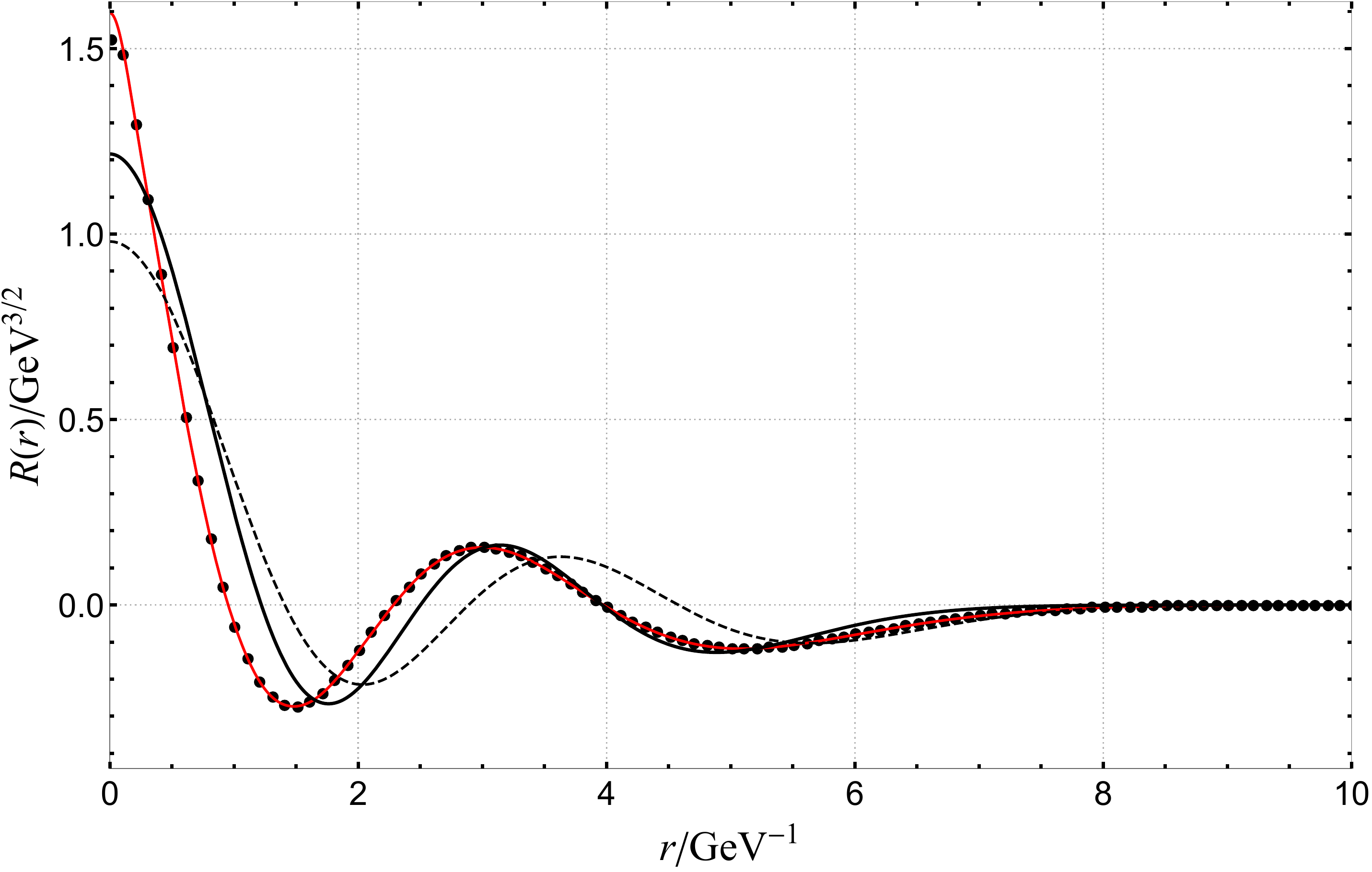}\\
  \caption{Comparison of $\Upsilon(4S)$'s spatial wave function.
   Numerical values and GEM fit are denoted by black dots and red solid curve, respectively.
   Black dashed and solid curve represent single SHO approximation by matching $\langle r\rangle$ and maximizing wave function overlap, respectively.
   }
\label{WaveCompare}
\end{figure}

\begin{figure}[h]
  \centering
  \includegraphics[width=0.8\textwidth]{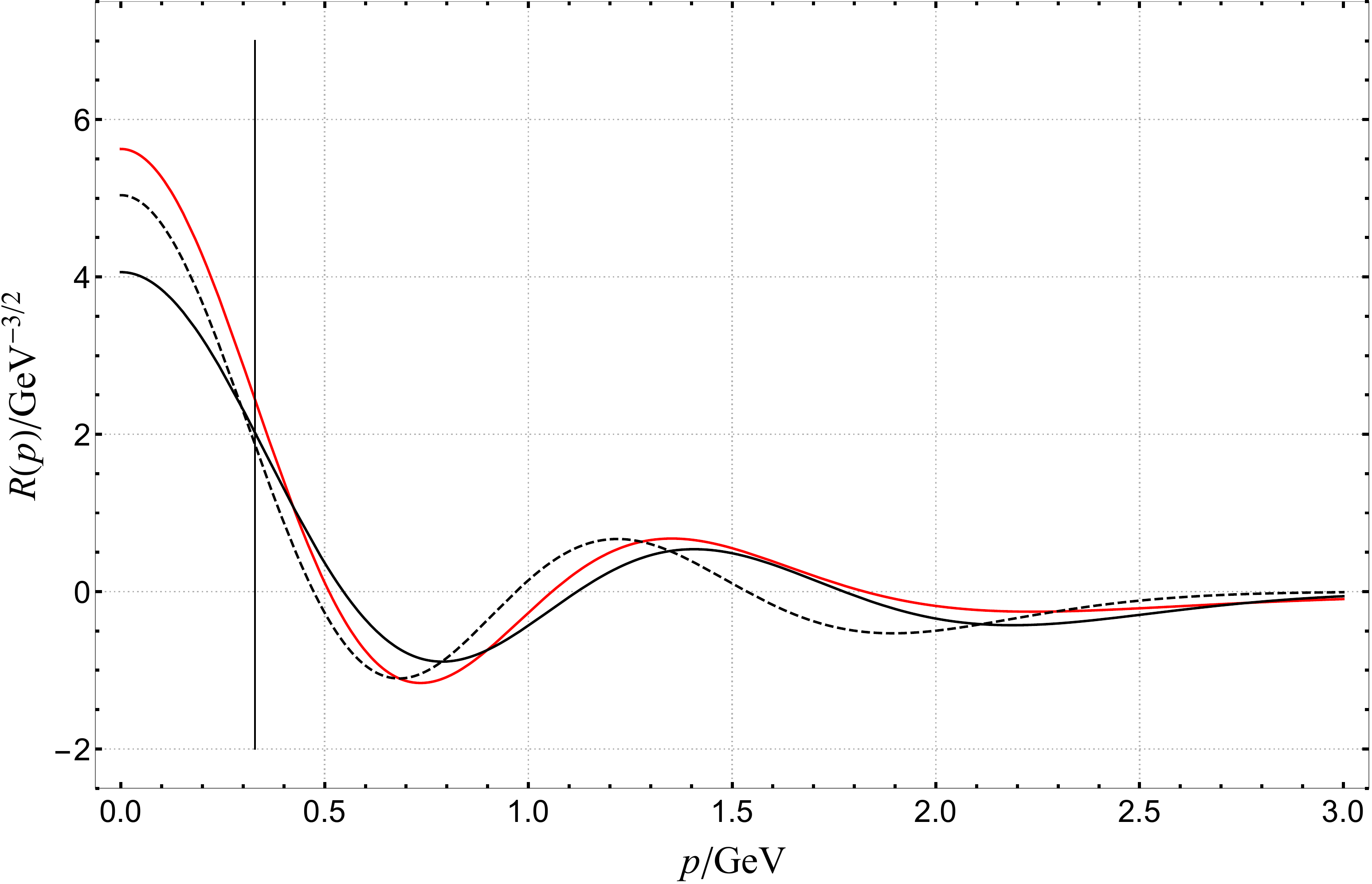}\\
  \caption{Comparison of $\Upsilon(4S)$ wave function in momentum space.
   GEM fit, single SHO approximation by matching $\langle r\rangle$ and maximizing wave function overlap
   are denoted by red solid curve, black dashed curve and black solid curve, respectively.
  Decay momentum for $B\bar{B}$ is shown as the vertical line.
 }
\label{MWaveCompare}
\end{figure}

To improve the accuracy, people also expand the true wave function in terms of SHOs (see e.g. Refs.~\cite{Blundell:1996as,Danilkin:2009hr}).
As a consequence, one will get fairly complicated analytic expression for highly excited states.
For example, expression~(2.12) in Ref.~\cite{Blundell:1996as}.
Due to the highly oscillated behavior of the excited SHOs,
one would need a large number of SHOs to achieve an ideal precision.

We think that it is necessary to fully respect the wave function and make a precise calculation of the transition amplitude,
which is shown to be essential for the states near threshold.
In this work, we get both the analytic expression and the high precision
by using the Gaussian expansion method (GEM) proposed by Hiyama \emph{et al}.~\cite{Hiyama:2003cu}.
This method has the observation that bound state's wave function can be expanded in Gaussian bases as the following
\be
\psi_{NLM}(r)=\left(\sum_{i=1}^n c_i \beta_i^{L+\frac{3}{2}}e^{-\frac{1}{2}\beta_i^2 r^2}  r^L \right)Y_L^M(\theta,\varphi),
\ee
where $\beta_i$s and $c_i$s denote oscillator parameters and corresponding coefficients, respectively.
$n$ is the number of Gaussian basis.
In this work, $n=5\sim20$ for initial states and 5 for $B$ mesons,
and $\beta_i$s lie in the range of $0.1\sim5$GeV.

Compared with the SHO basis, the Gaussian basis is no longer orthogonal.
So a little trick is used to speed up the fitting procedure.
As explained in Ref~\cite{Hiyama:2003cu},
$\beta_i$s are set to be a geometric series.
Instead of increasing the number of SHO basis for fixed $\beta$,
GEM can both increase the bases number and change $\beta$ to improve the fit.
As shown in Fig.~\ref{WaveCompare}, 
for the spatial wave function of $\Upsilon(4S)$, the quality of GEM fitting is quite impressive. 
In momentum space, the wave functions of different fitting methods are shown in Fig.~\ref{MWaveCompare}. 
Note that the overall resemblance of wave functions does not indicate a small deviation of decay width or mass shift (see Tab.~\ref{decayTab} and Fig.~\ref{shiftFig}).

Because the $\3P0$ model calculation is easily done in momentum space,
we need to make a Fourier transformation of the position space wave function.
One benefit of the SHO wave function is that it is invariant after Fourier transform
apart from the substitution $\beta\to 1/\beta$.
Since Gaussian basis is ground state SHO wave function, it naturally keeps this property.
That means, after fitting position space wave functions,
we can rebuild the momentum space wave functions by
\be
\psi_{NLM}(p)=\left(\sum_{i=1}^n c_i \beta_i^{-(L+\frac{3}{2})}e^{-\frac{p^2}{2\beta_i^2}}  p^L \right)Y_L^M(\theta,\varphi).
\ee
What makes GEM simple is that there are a minimum number of polynomials in the integration
which simplifies the expression from the beginning.
GEM is quite universal and it is not limited to the wave functions obtained by solving the nonrelativistic Schr\"{o}dinger equation.

To work out the analytic expression,
one has to deal with the associated Laguerre polynomials if SHO wave functions are involved
and also the sophisticated angular integration.
Although in Ref.~\cite{Roberts1992}, Roberts and B. Silvestre-Brac show us the general method to do the integration,
and there are analytical expressions~\cite{Blundell:1996as},
however, these expressions are quite lengthy and they only apply to $\3P0$ model.

The complexity can be bypassed if we transfer the spherical harmonics into Cartesian form~\cite{Schlegel:1995} from the beginning.
After this transformation, the general form of the integration can be compactly expressed as
\be
\iiint dp_x dp_y dp_z \exp(-p^\mathrm{T}.A.p-B.p-C) f(p_x,p_y,p_z),\label{GaussianIntegration}
\ee
where $p,A,B$ and $C$ in the exponent denote $(p_x,p_y,p_z)^\mathrm{T}$, $3\times 3$ real symmetric matrix,
three-vector and constant, respectively.
$f(p_x,p_y,p_z)$ is nothing but polynomial, so Eq.~(\ref{GaussianIntegration}) is a standard form of Gaussian integration
which can be easily done even manually.

After the integration is done, we can easily transform it back to spherical basis by the substitution
$P_{x}=P \sin\theta \cos\varphi, P_{y}=P \sin\theta \sin\varphi, P_{z}=P \cos\theta$.
Another benefit of this transformation is that it can easily handle much more complicated  polynomials $f(p_x,p_y,p_z)$
which may show up in other quark pair creation models.

\section{Result and Discussion}\label{resultSec}
\subsection{Mass Shift and Open Channel Strong Decay}
From Tab.~\ref{massTabTotal} and~\ref{massTabChannel},
one can find that the mass shifts are generally same between GEM and SHOs with a few exceptions for near threshold states.
The mass shifts in a same multiplet are also almost equal
simply because their wave functions are identical and their bare masses are approximately equal.
This conclusion is consistent with the loop theorem in Ref.~\cite{Barnes:2007xu}.

For the states below threshold, $\Delta M$s are all negative,
and closer to the threshold gets more deduction of the mass.
With GEM, this conclusion is true even for states slightly above threshold.
This conclusion differs with Refs.~\cite{Liu:2011yp,Ferretti:2013vua}, where SHO are used to calculate the mass shift.
Take $h_b$ family as an example, our mass shift grows with the mass going higher no matter whether we use GEM or SHO,
however, in Ref.~\cite{Liu:2011yp} and Ref.~\cite{Ferretti:2013vua},
the largest mass shift happens to $h_b(1P)$ and $h_b(2P)$, respectively.

\begin{figure}[H]
  \centering
  \includegraphics[width=\textwidth]{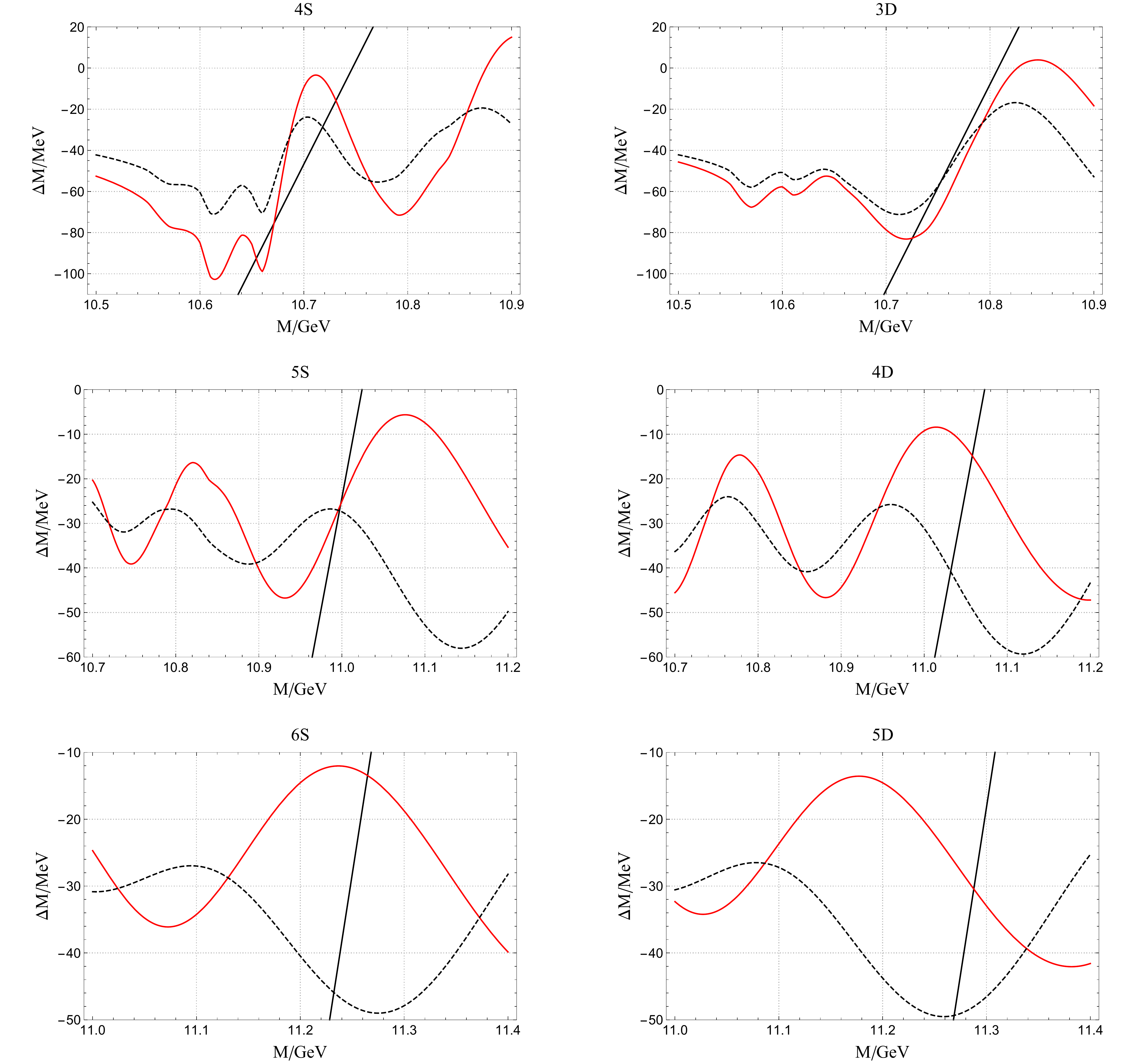}\\
  \caption{ The dependence of $\Delta M$ of
  $\Upsilon(nS)$ and $\Upsilon_1(nD)$ family on the mass of the initial states.
  GEM and SHO results are denoted by red solid curve and black dashed curve, respectively.
  $M-M_0$ is shown by the solid black line.
  One can read the $M$ (which are shown in Tab.~\ref{massTabTotal}) and corresponding $\Delta M$ from the intersection points of $M-M_0$ and $\Delta M$.
  }
\label{shiftFig}
\end{figure}

For states above threshold,
the mass shift behavior becomes complicated (see Fig.~\ref{shiftFig})
and it is not appropriate to draw the conclusion 
that the mass shift of state above the $B\bar{B}$  threshold is positive.
This conclusion is only true for asymptotically large mass, and in this case, $\Delta M \propto 1/M$.
We should also point out that this mathematical fact does not mean it will definitely happen.
The reason is that when mass becomes bigger, more $B$ meson channels will contribute,
and one cannot tell the sign of $\Delta M$ before summing all possible channels' contributions in Eq.~(\ref{mShift}).

\begin{figure}[H]
  \centering
  \includegraphics[width=\textwidth]{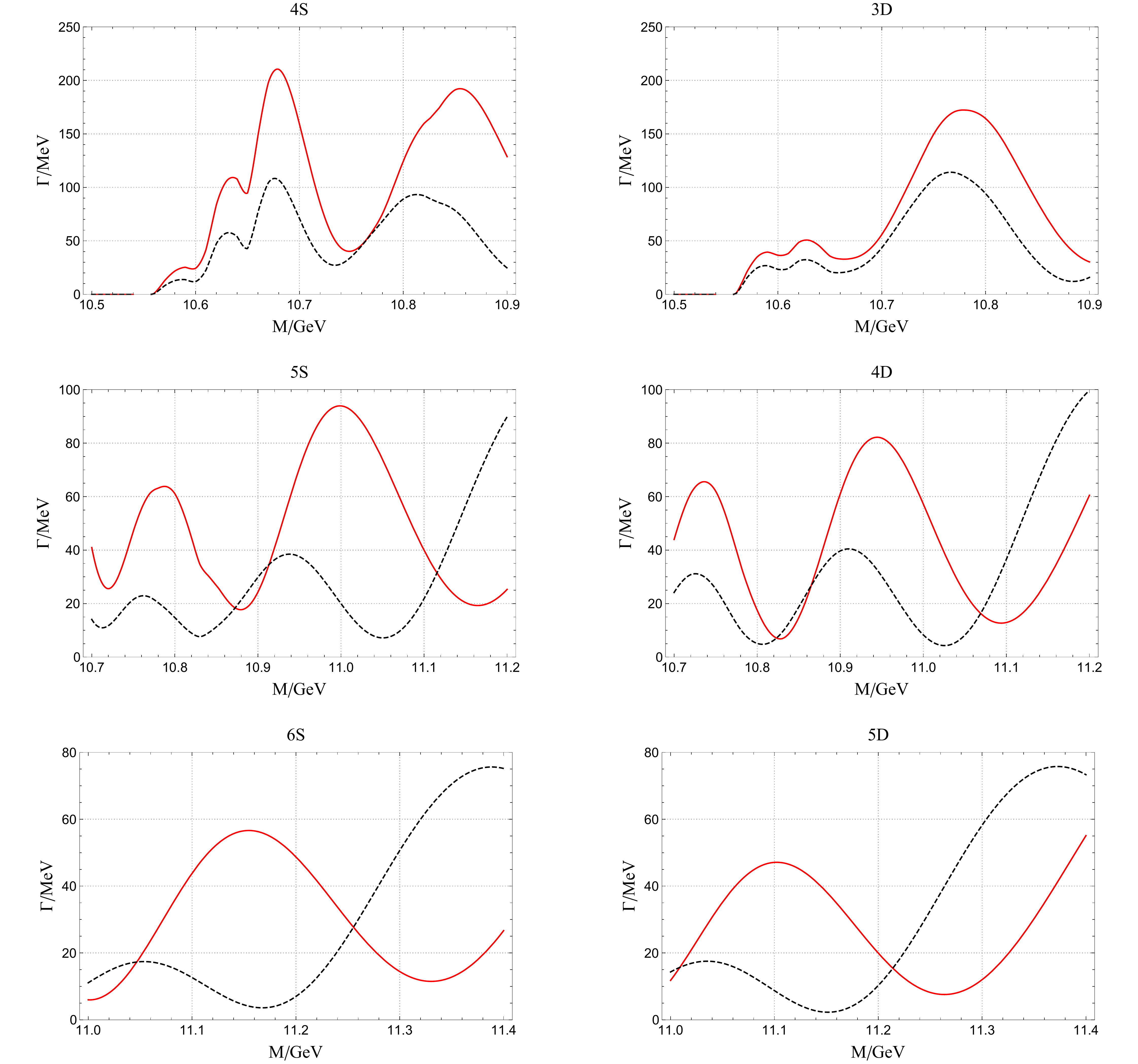}\\
  \caption{
  The dependence of the open channel strong decay widths of
  $\Upsilon(S)$ and $\Upsilon_1(D)$ family on the mass of the initial states.
  GEM and SHO results are denoted by red solid curve and black dashed curve, respectively.
  One can directly read the total decay width from this plot.
  }
\label{decayFig}
\end{figure}

In order to study this sensitivity,
we also plot the dependence on the initial state mass of $\Delta M$  and decay width
for the vector meson above threshold in Fig.~\ref{shiftFig} and Fig.~\ref{decayFig}, respectively.
As a concrete example, one can see this sensitivity by comparing $\Upsilon(4S)$ with $\Upsilon(6S)$.
Compared with $\Upsilon(4S)$, the wave function of $\Upsilon(6S)$ has more nodes,
however, basing on this fact one cannot conclude that the $\Delta M$'s behavior is more complicated.
The important reason is that the bare mass of $\Upsilon(6S)$ is also farther from threshold,
causing the average of the wave function overlap integration in Eq.~(\ref{overlap}).
Note also that absolute value of the mass shift of $\Upsilon(6S)$ calculated in SHO is larger than GEM,
however, in $\Upsilon(4S)$ case,
we have the opposite conclusion if we choose the lowest intersection point of $\Delta M$ and $M-M_0$.

From Tab.~\ref{massTabTotal},
one can find that the masses predicted in Ref.~\cite{Liu:2011yp} are generally closer to the experimental data.
However, we want to stress that the spectrum is an important but not the only criterion to judge whose parameters are better.
As shown in Fig.~\ref{eeDecayPlot},
the dielectric decay ratios $\Gamma_{ee}/\Gamma_{ee}(1S)$ calculated with the parameters given in Ref.~\cite{Liu:2011yp}
are generally smaller than experimental measurements before coupled-channel effects are taken into account.
As will be discussed in Sec.~\ref{eeRatioSec},
coupled-channel effects will suppress rather than enhance these ratios,
so their parameters are difficult to explain the dielectric decays of vector mesons despite of their success in the spectrum.

\begin{table}[H]\small
 \renewcommand\arraystretch{1.4}
  \centering
\begin{tabular}{c|c|cccc|cccc|c}
  \hline\hline
 States
 &$M_0$
 &\multicolumn{4}{c|}{$-\Delta M$}
 &\multicolumn{4}{c|}{$M_\text{theory}$}
 &$M_\text{exp}$
  \\
 \hline
 &
 &GEM&SHO&Ref.~\cite{Liu:2011yp} & Ref.~\cite{Ferretti:2013vua}
 &GEM&SHO&Ref.~\cite{Liu:2011yp} & Ref.~\cite{Ferretti:2013vua}
 &
 \\
 \hline
 $\eta_b(1{}^1S_0)$ & 9416.5 & 22.0 & 22.0 & 55.5 & 64 & 9394.5 & 9394.5 & 9391.8 & 9391 & 9398.0 \\
 $\eta_b(2{}^1S_0)$ & 10024.2 & 42.4 & 41.5 & 66.2 & 101 & 9981.8 & 9982.6 & 10004.9 & 9980 & 9999.0 \\
 $\eta_b(3{}^1S_0)$ & 10410.0 & 57.4 & 51.4 & 66.4 & 129 & 10352.7 & 10358.6 & 10337.9 & 10338 & -- \\
  \hline
 $\Upsilon (1{}^3S_1)$ & 9482.0 & 22.8 & 22.8 & 58.2 & 69 & 9459.2 & 9459.2 & 9460.3 & 9489 & 9460.3 \\
 $\Upsilon (2{}^3S_1)$ & 10054.9 & 43.8 & 42.8 & 68.0 & 108 & 10011.2 & 10012.1 & 10026.2 & 10022 & 10023.3 \\
 $\Upsilon (3{}^3S_1)$ & 10433.4 & 60.0 & 53.5 & 68.2 & 146 & 10373.4 & 10379.9 & 10351.9 & 10358 & 10355.2 \\
 $\Upsilon (4{}^3S_1)$ & 10746.7 & 92.6 & 28.7 & 76.3 & -- & 10654.2 & 10718.0 & 10602.7 & -- & 10579.4 \\
 $\Upsilon (5{}^3S_1)$ & 11024.3 & 25.7 & 27.2 & 84.2 & -- & 10998.6 & 10997.1 & 10819.9 & -- & 10876.0 \\
 $\Upsilon (6{}^3S_1)$ & 11278.2 & 13.5 & 45.9 & 85.5 & -- & 11264.8 & 11232.3 & 11022.6 & -- & 11019.0 \\
  \hline
 $\Upsilon_1(1{}^3D_1)$ & 10181.9 & 46.1 & 49.1 & 96.8 & 159 & 10135.7 & 10132.8 & 10138.1 & 10112 & -- \\
 $\Upsilon_1(2{}^3D_1)$ & 10515.9 & 62.3 & 62.0 & 88.4 & -- & 10453.6 & 10453.9 & 10420.4 & -- & -- \\
 $\Upsilon_1(3{}^3D_1)$ & 10807.9 & 82.6 & 55.7 & 93.4 & -- & 10725.2 & 10752.2 & 10650.9 & -- & -- \\
 $\Upsilon_1(4{}^3D_1)$ & 11072.8 & 14.8 & 40.8 & -- & -- & 11057.9 & 11031.9 & -- & -- & -- \\
 $\Upsilon_1(5{}^3D_1)$ & 11318.2 & 30.4 & 49.3 & -- & -- & 11287.8 & 11268.9 & -- & -- & -- \\
  \hline
 $h_b(1{}^1P_1)$ & 9921.7 & 35.8 & 37.3 & 85.7 & 115 & 9885.9 & 9884.4 & 9915.5 & 9885 & 9899.3 \\
 $h_b(2{}^1P_1)$ & 10315.4 & 53.1 & 52.7 & 78.8 & 146 & 10262.3 & 10262.7 & 10259.1 & 10247 & 10259.8 \\
 $h_b(3{}^1P_1)$ & 10637.9 & 77.9 & 69.4 & 79.8 & 114 & 10560.1 & 10568.5 & 10523.2 & 10591 & -- \\
  \hline
 $\chi_{b0}(1{}^3P_0)$ & 9886.1 & 34.6 & 36.0 & 81.8 & 108 & 9851.4 & 9850.0 & 9875.3 & 9879 & 9859.4 \\
 $\chi_{b0}(2{}^3P_0)$ & 10284.2 & 50.9 & 50.6 & 75.0 & 137 & 10233.4 & 10233.6 & 10227.9 & 10226 & 10232.5 \\
 $\chi_{b0}(3{}^3P_0)$ & 10608.7 & 76.1 & 68.6 & 75.7 & 186 & 10532.6 & 10540.2 & 10495.9 & 10495 & -- \\
 \hline
 $\chi_{b1}(1{}^3P_1)$ & 9915.4 & 35.5 & 37.0 & 84.8 & 114 & 9879.9 & 9878.4 & 9906.8 & 9879 & 9892.8 \\
 $\chi_{b1}(2{}^3P_1)$ & 10310.0 & 52.6 & 52.3 & 77.9 & 144 & 10257.4 & 10257.7 & 10252.4 & 10244 & 10255.5 \\
 $\chi_{b1}(3{}^3P_1)$ & 10632.9 & 77.4 & 69.0 & 78.8 & 121 & 10555.6 & 10563.9 & 10517.3 & 10580 & 10512.1 \\
  \hline
 $\chi_{b2}(1{}^3P_2)$ & 9934.9 & 36.4 & 37.8 & 87.3 & 117 & 9898.5 & 9897.1 & 9929.6 & 9900 & 9912.21 \\
 $\chi_{b2}(2{}^3P_2)$ & 10327.6 & 54.1 & 53.7 & 80.4 & 149 & 10273.5 & 10273.9 & 10270.1 & 10257 & 10268.7 \\
 $\chi_{b2}(3{}^3P_2)$ & 10649.8 & 82.2 & 73.6 & 82.1 & 138 & 10567.6 & 10576.2 & 10532.4 & 10578 & -- \\
  \hline
 $\Upsilon_2(1{}^3D_2)$ & 10187.8 & 46.6 & 49.5 & 97.7 & 161 & 10141.2 & 10138.3 & 10144.6 & 10121 & 10163.7 \\
\hline \hline
\end{tabular}
\caption{Total mass shift (in MeV) induced by coupled-channel effects.
$M_0$ denotes the bare mass of the Cornell potential whose parameters are shown in Tab.~\ref{paraTab}.
$M_{\text{theory}}$ is the mass after considering coupled-channel effects.
The last two columns of $-\Delta M$ and $M_{\text{theory}}$ are taken from
Ref.~\cite{Liu:2011yp} and Ref.~\cite{Ferretti:2013vua}, whose parameters are different from ours.
$M_{\text{exp}}$ denotes the experimental measured value.
For simplicity, the experimentally measured $\Upsilon(10580),\Upsilon(10860)$, and $\Upsilon(11020)$ are
assumed to be $\Upsilon(4S),\Upsilon(5S)$, and $\Upsilon(6S)$, respectively.
``--'' represents the corresponding value is not available.
}
\label{massTabTotal}
\end{table}

\begin{table}[H]\small
 \renewcommand\arraystretch{1.4}
  \centering
\begin{tabular}{c|cc|cc|cc|cc|cc|cc}
  \hline\hline
 States
 &\multicolumn{2}{c|}{$B\bar{B}$}
 &\multicolumn{2}{c|}{$B\bar{B}^*+h.c.$}
 &\multicolumn{2}{c|}{$B^*\bar{B}^*$}
 &\multicolumn{2}{c|}{$B_s\bar{B}_s$}
 &\multicolumn{2}{c|}{$B_s\bar{B}_s^*+h.c.$}
 &\multicolumn{2}{c}{$B_s^*\bar{B}^*_s$}
  \\
 \hline
 &GEM&SHO
 &GEM&SHO
 &GEM&SHO
 &GEM&SHO
 &GEM&SHO
 &GEM&SHO
 \\
 \hline
 $\eta_b(1{}^1S_0)$ & 0 & 0 & 7.8 & 7.8 & 7.6 & 7.6 & 0 & 0 & 3.3 & 3.3 & 3.3 & 3.3 \\
 $\eta_b(2{}^1S_0)$ & 0 & 0 & 16.5 & 16.1 & 15.7 & 15.4 & 0 & 0 & 5.2 & 5.1 & 5.0 & 4.9 \\
 $\eta_b(3{}^1S_0)$ & 0 & 0 & 24.5 & 21.8 & 22.3 & 20.0 & 0 & 0 & 5.4 & 4.9 & 5.1 & 4.7 \\
 \hline
 $\Upsilon (1{}^3S_1)$ & 1.4 & 1.4 & 5.4 & 5.4 & 9.2 & 9.2 & 0.6 & 0.6 & 2.3 & 2.3 & 3.9 & 3.9 \\
 $\Upsilon (2{}^3S_1)$ & 3.0 & 2.9 & 11.4 & 11.1 & 18.9 & 18.5 & 0.9 & 0.9 & 3.5 & 3.5 & 5.9 & 5.9 \\
 $\Upsilon (3{}^3S_1)$ & 4.8 & 4.2 & 17.2 & 15.2 & 27.1 & 24.3 & 1.0 & 0.9 & 3.7 & 3.4 & 6.1 & 5.6 \\
 $\Upsilon (4{}^3S_1)$ & -0.7 & 3.7 & -2.4 & 16.0 & 85.4 & -0.6 & 1.0 & 1.0 & 3.6 & 3.3 & 5.7 & 5.2 \\
 $\Upsilon (5{}^3S_1)$ & -0.5 & 2.8 & 2.8 & 6.8 & 17.8 & 10.1 & 0.8 & 0.7 & 1.7 & 2.7 & 3.1 & 4.0 \\
 $\Upsilon (6{}^3S_1)$ & 1.5 & 3.5 & 2.4 & 14.2 & 1.5 & 21.2 & 0.6 & 0.6 & 2.8 & 2.3 & 4.7 & 4.1 \\
 \hline
 $\Upsilon_1(1{}^3D_1)$ & 4.0 & 4.3 & 3.7 & 4.0 & 27.8 & 29.2 & 1.0 & 1.1 & 1.0 & 1.1 & 8.7 & 9.3 \\
 $\Upsilon_1(2{}^3D_1)$ & 9.0 & 8.7 & 7.4 & 7.3 & 35.2 & 35.2 & 1.4 & 1.4 & 1.2 & 1.3 & 8.1 & 8.2 \\
 $\Upsilon_1(3{}^3D_1)$ & 7.5 & 2.3 & 6.1 & 7.4 & 57.8 & 35.4 & 2.3 & 2.0 & 1.5 & 1.4 & 7.4 & 7.2 \\
 $\Upsilon_1(4{}^3D_1)$ & 0.1 & 6.2 & -1.6 & 3.6 & 6.8 & 22.5 & 1.2 & 1.0 & 1.3 & 1.1 & 7.0 & 6.4 \\
 $\Upsilon_1(5{}^3D_1)$ & 3.3 & 5.6 & 1.8 & 6.5 & 19.1 & 30.0 & 0.5 & 0.9 & 0.8 & 0.8 & 5.0 & 5.5 \\
 \hline
 $h_b(1{}^1P_1)$ & 0 & 0 & 13.5 & 14.0 & 13.0 & 13.4 & 0 & 0 & 4.8 & 5.0 & 4.6 & 4.8 \\
 $h_b(2{}^1P_1)$ & 0 & 0 & 21.9 & 21.6 & 20.3 & 20.2 & 0 & 0 & 5.6 & 5.6 & 5.3 & 5.3 \\
 $h_b(3{}^1P_1)$ & 0 & 0 & 38.0 & 33.5 & 29.5 & 26.3 & 0 & 0 & 5.4 & 5.0 & 5.0 & 4.6 \\
 \hline
 $\chi_{b0}(1{}^3P_0)$ & 4.1 & 4.3 & 0 & 0 & 21.4 & 22.2 & 1.3 & 1.4 & 0 & 0 & 7.8 & 8.1 \\
 $\chi_{b0}(2{}^3P_0)$ & 9.3 & 9.0 & 0 & 0 & 31.1 & 31.0 & 2.1 & 2.1 & 0 & 0 & 8.4 & 8.5 \\
 $\chi_{b0}(3{}^3P_0)$ & 25.5 & 22.4 & 0 & 0 & 40.7 & 36.9 & 2.3 & 2.0 & 0 & 0 & 7.6 & 7.2 \\
  \hline
 $\chi_{b1}(1{}^3P_1)$ & 0 & 0 & 10.8 & 11.2 & 15.5 & 16.0 & 0 & 0 & 3.7 & 3.9 & 5.6 & 5.9 \\
 $\chi_{b1}(2{}^3P_1)$ & 0 & 0 & 19.7 & 19.4 & 22.1 & 22.0 & 0 & 0 & 4.8 & 4.8 & 6.0 & 6.0 \\
 $\chi_{b1}(3{}^3P_1)$ & 0 & 0 & 37.4 & 32.6 & 29.7 & 26.9 & 0 & 0 & 4.8 & 4.4 & 5.4 & 5.2 \\
  \hline
 $\chi_{b2}(1{}^3P_2)$ & 3.4 & 3.5 & 9.8 & 10.1 & 13.6 & 14.2 & 1.2 & 1.3 & 3.5 & 3.7 & 4.7 & 5.0 \\
 $\chi_{b2}(2{}^3P_2)$ & 5.3 & 5.2 & 14.6 & 14.4 & 23.2 & 23.0 & 1.3 & 1.3 & 3.8 & 3.8 & 5.8 & 5.9 \\
 $\chi_{b2}(3{}^3P_2)$ & 12.3 & 11.2 & 23.3 & 20.7 & 36.2 & 32.0 & 1.3 & 1.2 & 3.6 & 3.3 & 5.6 & 5.2 \\
 \hline
 $\Upsilon_2(1{}^3D_2)$ & 0 & 0 & 16.0 & 17.0 & 19.8 & 20.9 & 0 & 0 & 4.6 & 5.0 & 6.2 & 6.6 \\
\hline \hline
\end{tabular}
\caption{The mass shift (in MeV) of every coupled channel.
Coupled-channel induced $S-D$ mixing is not considered in this table.
0 represents that the contributions of some channels are forbidden.
For simplicity, an overall negative sign has been omitted for all the channels.
Note that for a few channels, mass shifts are positive.}
\label{massTabChannel}
\end{table}

The complicated structure of the mass shift of $\Upsilon (4S)$ needs further discussion.
Even though the curves of GEM and SHO share some common features,
the small difference is sufficient to generate a large discrepancy of the mass shift.
Another interesting feature of this plot is that GEM has three solutions,
implying that more resonances may pop up compared with potential model prediction.

This sensitivity can also be seen by the decay width behavior in Fig.~\ref{decayFig}.
Over a large energy range $10.58\sim 10.73$ GeV,
the decay width of $\Upsilon(4S)$ calculated by GEM can be around two times as large as SHO.
Our decay width plot of $\Upsilon(4S)$ also shares some resemblance with Fig.2 in Ref.~\cite{Ono:1980js},
where the prediction of SHO is not calculated.

The deviations of mass shift and decay width tell us that
it is necessary to adopt the realistic wave functions other than SHO approximation in the coupled channel calculation.

Of course, one may argue that, since the bare mass is 140MeV heavier
than the experimental measurement,
if the bare mass is tuned closer to the $B\bar{B}$ threshold,
the difference between GEM and SHO would be small and we will get one solution.
However, we want to stress that the bare mass is directly related to the wave function,
in the case where we have a smaller bare mass, the wave function will also be different,
thus causing different mass shift behavior.
This sensitivity also reminds us that taking only one step approximation
in the recursive method to solve Eq.~(\ref{mShift}) may cause a large error,
so an accurate treatment of wave function and a precise method to solve Eq.~(\ref{mShift})
are essential for near threshold states.

\begin{table} [H]\footnotesize
\renewcommand\arraystretch{1.3}
  \centering
\begin{tabular}{c|cc|cc|cc|cc|cc|cc|cc}
  \hline\hline
 States
 &\multicolumn{2}{c|}{$B\bar{B}$}
 &\multicolumn{2}{c|}{$B\bar{B}^*+h.c.$}
 &\multicolumn{2}{c|}{$B^*\bar{B}^*$}
 &\multicolumn{2}{c|}{$B_s\bar{B}_s$}
 &\multicolumn{2}{c|}{$B_s\bar{B}_s^*+h.c.$}
 &\multicolumn{2}{c|}{$B_s^*\bar{B}^*_s$}
 &\multicolumn{2}{c}{$P_{b\bar{b}}$}
 \\ \hline
 &GEM&SHO
 &GEM&SHO
 &GEM&SHO
 &GEM&SHO
 &GEM&SHO
 &GEM&SHO
 &GEM&SHO
 \\ \hline
 $\eta_b(1S)$ & 0 & 0 & 0.45 & 0.44 & 0.43 & 0.42 & 0 & 0 & 0.17 & 0.17 & 0.16 & 0.16 & 98.79 & 98.81 \\
 $\eta_b(2S)$ & 0 & 0 & 1.81 & 1.65 & 1.62 & 1.49 & 0 & 0 & 0.43 & 0.4 & 0.4 & 0.37 & 95.74 & 96.08 \\
 $\eta_b(3S)$ & 0 & 0 & 5.01 & 4.02 & 3.98 & 3.24 & 0 & 0 & 0.63 & 0.51 & 0.55 & 0.45 & 89.83 & 91.78 \\
 \hline
 $\Upsilon(1S)$ & 0.09 & 0.08 & 0.33 & 0.32 & 0.54 & 0.53 & 0.03 & 0.03 & 0.12 & 0.12 & 0.2 & 0.2 & 98.69 & 98.72 \\
 $\Upsilon(2S)$ & 0.37 & 0.33 & 1.29 & 1.18 & 2.02 & 1.86 & 0.08 & 0.08 & 0.31 & 0.28 & 0.49 & 0.45 & 95.44 & 95.82 \\
 $\Upsilon(3S)$ & 1.25 & 0.99 & 3.71 & 2.98 & 5.07 & 4.12 & 0.13 & 0.1 & 0.45 & 0.36 & 0.67 & 0.55 & 88.71 & 90.89 \\
  \hline
 $\Upsilon_1(1D)$ & 0.65 & 0.69 & 0.55 & 0.59 & 2.84 & 2.94 & 0.12 & 0.13 & 0.1 & 0.12 & 0.71 & 0.74 & 95.03 & 94.79 \\
 $\Upsilon_1(2D)$ & 3.76 & 3.36 & 2.36 & 2.15 & 6.21 & 5.84 & 0.24 & 0.22 & 0.19 & 0.18 & 0.83 & 0.8 & 86.41 & 87.45 \\
  \hline
 $h_b(1P)$ & 0 & 0 & 1.22 & 1.24 & 1.12 & 1.14 & 0 & 0 & 0.35 & 0.37 & 0.33 & 0.34 & 96.99 & 96.91 \\
 $h_b(2P)$ & 0 & 0 & 3.51 & 3.24 & 2.96 & 2.76 & 0 & 0 & 0.59 & 0.56 & 0.52 & 0.5 & 92.43 & 92.94 \\
 $h_b(3P)$ & 0 & 0 & 19.75 & 18.19 & 9.04 & 7.7 & 0 & 0 & 0.67 & 0.54 & 0.54 & 0.45 & 70.0 & 73.12 \\
  \hline
 $\chi_{b0}(1P)$ & 0.45 & 0.46 & 0 & 0 & 1.74 & 1.77 & 0.11 & 0.12 & 0 & 0 & 0.52 & 0.55 & 97.18 & 97.1 \\
 $\chi_{b0}(2P)$ & 1.85 & 1.68 & 0 & 0 & 4.13 & 3.88 & 0.26 & 0.25 & 0 & 0 & 0.77 & 0.75 & 92.98 & 93.45 \\
 $\chi_{b0}(3P)$ & 34.08 & 38.84 & 0 & 0 & 8.07 & 6.21 & 0.31 & 0.22 & 0 & 0 & 0.62 & 0.48 & 56.92 & 54.26 \\
  \hline
 $\chi_{b1}(1P)$ & 0 & 0 & 1.03 & 1.06 & 1.27 & 1.29 & 0 & 0 & 0.28 & 0.3 & 0.38 & 0.4 & 97.03 & 96.95 \\
 $\chi_{b1}(2P)$ & 0 & 0 & 3.38 & 3.11 & 3.0 & 2.81 & 0 & 0 & 0.53 & 0.51 & 0.56 & 0.54 & 92.53 & 93.04 \\
 $\chi_{b1}(3P)$ & 0 & 0 & 21.9 & 20.1 & 7.54 & 6.44 & 0 & 0 & 0.64 & 0.51 & 0.54 & 0.46 & 69.38 & 72.5 \\
  \hline
 $\chi_{b2}(1P)$ & 0.31 & 0.31 & 0.85 & 0.87 & 1.24 & 1.27 & 0.09 & 0.09 & 0.25 & 0.26 & 0.35 & 0.37 & 96.91 & 96.83 \\
 $\chi_{b2}(2P)$ & 0.89 & 0.82 & 2.23 & 2.06 & 3.62 & 3.36 & 0.15 & 0.14 & 0.39 & 0.37 & 0.6 & 0.58 & 92.13 & 92.68 \\
  \hline
 $\Upsilon_2(1D)$ & 0 & 0 & 2.0 & 2.1 & 2.08 & 2.16 & 0 & 0 & 0.44 & 0.47 & 0.51 & 0.54 & 94.98 & 94.74 \\
  \hline \hline
\end{tabular}
\caption{Probabilities of every coupled channel and $b\bar{b}$ component for states below threshold.
The effect of $S-D$ mixing is not considered in this table.
0 represents that the contributions of some channels are forbidden.
The overall \% has been omitted for simplicity.
Note that despite $m(h_b(3P))$ calculated with GEM and SHO and $m(\chi_{b1}(3P))$ calculated with SHO are above $B\bar{B}$ threshold,
however, they do not coupled to $B\bar{B}$ and their masses are still small than $B\bar{B}^*$,
so the probabilities of $B$ meson continuum are well defined.
}
\label{ratioTab}
\end{table}

If $\Upsilon(10580),\Upsilon(10860)$, and $ \Upsilon(11020)$ are treated to be pure $S$ or $D$ wave,
we get the open channel decay width shown in Tab.~\ref{decayTab}.
It is worthy to note that this assumption is oversimplified, so the absolute value cannot be treated too seriously.

\begin{table}[H]\scriptsize
\renewcommand\arraystretch{1.5}
  \centering
\begin{tabular}{c|cc|cc|cc|cc|cc|cc|cc|cccccccccc}
  \hline\hline
 State
 &\multicolumn{2}{c|}{$B\bar{B}$}
 &\multicolumn{2}{c|}{$B\bar{B}^*+h.c.$}
 &\multicolumn{2}{c|}{$B^*\bar{B}^*$}
 &\multicolumn{2}{c|}{$B_s\bar{B}_s$}
 &\multicolumn{2}{c|}{$B_s\bar{B}_s^*+h.c.$}
 &\multicolumn{2}{c|}{$B_s^*\bar{B}^*_s$}
 &\multicolumn{2}{c|}{$\Gamma_\text{theory}$}
 &$\Gamma_\text{exp}$
  \\ \cline{2-15}
 &GEM&SHO
 &GEM&SHO
 &GEM&SHO
 &GEM&SHO
 &GEM&SHO
 &GEM&SHO
 &GEM&SHO
 &
 \\ \hline
$4S$ & 21.1 & 12.5 & 0 & 0 & 0 & 0 & 0 & 0 & 0 & 0 & 0 & 0 & 21.1 & 12.5 &\multirow{2}*{$20.5\pm2.5$}\\
$3D$ & 34.1 & 24.2 & 0 & 0 & 0 & 0 & 0 & 0 & 0 & 0 & 0 & 0 & 34.1 & 24.2 & \\
\hline
$5S$ & 5.1 & 3.5 & 4.8 & 11.1 & 1.9 & 4.1 & 0.9 & 0.2 & 0.6 & 0.4 & 4.5 & 0.5 & 17.9 & 19.7 &\multirow{2}*{$55\pm 28$}\\
$4D$ & 10.8 & 7.2 & 4.0 & 5.4 & 18.1 & 18.1 & 1.21 & 0.3 & 0.3 & 0.2 & 2.8 & 0.9 & 37.3 & 32.1& \\
 \hline
$6S$ & 2.9 & 1.3 & 3.4 & 6.4 & 0.1 & 6.5 & 0.3 & 0.0 & 1.0 & 0.1 & 0.2 & 0.2 & 7.8 & 14.5 & \multirow{2}*{$79\pm 16$}\\
$5D$ & 6.5 & 3.0 & 2.9 & 3.3 & 9.2 & 10.1 & 0.4 & 0.0 & 0.4 & 0.1 & 1.1 & 0.2 & 20.4 & 16.8& \\
 \hline \hline
\end{tabular}
\caption{Open channel strong decay widths (in MeV) of pure $S$ and $D$ wave vector bottomonia.
$\Upsilon(10580),\Upsilon(10860),\Upsilon(11020)$ are considered to be close to
$4S\sim3D,5S\sim4D,6S\sim5D$, respectively.
}
\label{decayTab}
\end{table}

\subsection{$S-D$ Mixing and Dielectric Decay}\label{eeRatioSec}

As explained in Sec.~\ref{modelSec} and sketched in Fig.~\ref{ccDiagram},
coupled-channel effects will also induce mixing among states with same $J^{PC}$.
In this paper, we focus on the mixing between $\Upsilon(S)$ and $\Upsilon_1(D)$ family.
The Cornell potential model tells us that
the mass splitting between $\Upsilon((n+1)S)$ and $\Upsilon_1(nD)$ is smaller than other configurations,
such as $\Upsilon(nS)$ and $\Upsilon((n+1)S)$ or $\Upsilon_1(nD)$ and $\Upsilon_1((n+1)D)$ states,
so its quite reasonable to assume that the mixing only happens to $\Upsilon((n+1)S)$ and $\Upsilon_1(nD)$.
The masses and corresponding mixing angles after considering $S-D$ mixing are listed in Tab.~\ref{mixTab}.

In Eq.~(\ref{SDMixing}), the overall phase before $c_S$ and $ c_D$ is nonphysical,
so we are free to set the phase of $c_D$ to be 0, i.e. $c_D\ge 0$.
Under this convention and the normalization condition Eq.~(\ref{SDMixingRatio}),
the ratio $c_S/c_D$ is adequate to fix the value of $c_S$ and $c_D$.
If the imaginary part of $\Delta M_f$ and $\Delta M_{SD}$ in Eq.~(\ref{SDMixing}) are neglected,
one would get real solutions both for $M$ and $c_S/c_D$.
After this approximation, one can deduce the mixing angles.
However, the definition of $c_S$ and $c_D$ does not exist for states above threshold~\cite{Kalashnikova:2005ui,Baru:2010ww}.
Despite of this difficulty, we follow Ref.~\cite{Heikkila:1983wd},
assuming that these open channels' contribution are neglected.
Under this assumption, quantities related to $S-D$ mixing are shown in Tab.~\ref{mixTab}.

\begin{table}[H]\scriptsize
\renewcommand\arraystretch{1.9}
  \centering
\begin{tabular}{c|c|cc|cc|cc|cc|cc}
  \hline\hline
 & &2S&1D &3S&2D &4S&3D &5S&4D &6S&5D\\ \cline{2-12}
 &$M_0$ &10.055 & 10.182 & 10.433 & 10.516 & 10.747 & 10.808 & 11.024 & 11.073 & 11.278 & 11.318 \\ \hline\hline
 \multirow{6}*{GEM}
 &$M_{\text{pure}}$             &10.011 & 10.136 & 10.373 & 10.454 & 10.654 & 10.725 & 10.999 & 11.058 & 11.265 & 11.288 \\
 \cline{2-12}
 &$M_{\text{comp}}$       &10.011 & 10.136 & 10.373 & 10.454 & \tbreak{c}{10.651\\ +0.047i} & \tbreak{c}{10.731\\ +0.032i} & \tbreak{c}{10.999\\ +0.047i} & \tbreak{c}{11.058\\ +0.01i} & \tbreak{c}{11.265\\ +0.012i} & \tbreak{c}{11.288\\ +0.005i} \\
 &$M_{\text{real}}$             &10.011 & 10.136 & 10.373 & 10.454 & 10.653 & 10.734 & 10.999 & 11.058 & 11.265 & 11.288 \\
 \cline{2-12}
 &$c_S/c_D(\text{comp})$ &5482 & 0.0 & 524 & -0.005 & \tbreak{c}{2.55\\ +2.63i} & \tbreak{c}{2.10\\ -1.16i} & \tbreak{c}{32.37\\ +12.65i} & \tbreak{c}{-0.03\\ -0.002i} & \tbreak{c}{10.55\\ -40.41i} & \tbreak{c}{-0.01\\ +0.02i} \\
 &$c_S/c_D(\text{real})$        &5482 & 0.0 & 524 & -0.005 & 6.19 & 0.97 & 41.8 & -0.03 & 77.2 & -0.005 \\
  \cline{2-12}
 &$\theta ^\circ$              &0.01 & 0.02 & 0.11 & 0.27 & 9.18 & 44.1 & 1.37 & 1.79 & 0.74 & 0.3 \\
\hline\hline
\multirow{6}*{SHO}
 &$M_{\text{pure}}$            &10.012 & 10.133 & 10.38 & 10.454 & 10.718 & 10.752 & 10.997 & 11.032 & 11.232 & 11.269 \\
 \cline{2-12}
 &$M_{\text{comp}}$     &10.012 & 10.133 & 10.38 & 10.454 & \tbreak{c}{10.716\\ +0.021i} & \tbreak{c}{10.754\\ +0.055i} & \tbreak{c}{10.997\\ +0.011i} & \tbreak{c}{11.032\\ +0.002i} & \tbreak{c}{11.232\\ +0.009i} & \tbreak{c}{11.269\\ +0.021i} \\
 &$M_{\text{real}}$            &10.012 & 10.133 & 10.38 & 10.454 & 10.717 & 10.754 & 10.997 & 11.032 & 11.232 & 11.269 \\
 \cline{2-12}
 &$c_S/c_D(\text{comp})$    &-5750 & 0.0 & -584 & 0.004 & \tbreak{c}{0.52\\ +2.73i} & \tbreak{c}{-0.046\\ +0.065i} & \tbreak{c}{-19.32\\ +20.87i} & \tbreak{c}{0.007\\ -0.037i} & \tbreak{c}{37.81\\ +23.61i} & \tbreak{c}{-0.021\\ +0.004i} \\
 &$c_S/c_D(\text{real})$         & -5750 & 0.0 & -584 & 0.004 & 3.58 & -0.085 & -36.6 & 0.005 & 49.9 & -0.02 \\
 \cline{2-12}
 &$\theta ^\circ$              &0.01 & 0.02 & 0.1 & 0.24 & 15.6 & 4.8 & 1.56 & 0.28 & 1.15 & 1.21 \\
 \hline \hline
\end{tabular}
\caption{Mixing between $\Upsilon((n+1)S)$ and $\Upsilon_1 (nD)$ calculated with GEM and SHO.
The unit of mass is GeV.
$M_0$ is the bare mass calculated by potential model.
$M_\text{pure}$ is taken from column 7 in Tab.~\ref{massTabTotal}, where $S-D$ mixing is not considered.
$M_\text{comp}$ and $M_\text{real}$ both denote the masses after $S-D$ mixing,
the difference is that the latter is the solution when one neglects the imaginary part of Eq.~(\ref{SDMixing}),
while the former is the precise solution of Eq.~(\ref{SDMixing}),
and its imaginary part equals to the one half of the decay width.
$c_S/c_D(\text{comp})$ and $c_S/c_D(\text{real})$ denote the ratio $c_S/c_D$ corresponding to $M_\text{comp}$ and $M_\text{real}$, respectively.
Note that after 3-digits approximation of the mass,
$M_\text{pure}$ may be the same with $M_{\text{real}}$, which is in fact different,
and because $c_S/c_D$ is too small to show for $1D$ case, we use 0.0 instead.
}
\label{mixTab}
\end{table}

From Tab.~\ref{mixTab}, we found that the masses barely change
after considering $S-D$ mixing for states below threshold,
indicating that the mixing angles are approximately 0.
So it is reasonable to treat $\Upsilon(1S),\Upsilon(2S)$, and $\Upsilon(3S)$ as pure $S$ wave states.
This conclusion also agrees with the loop theorem in Ref.~\cite{Barnes:2007xu}.
From Eq.~(\ref{eeFormula}),
we also learn that the dielectric decay of $\Upsilon_1(D)$ is suppressed by the $b$ quark mass $m_b$,
so the small mixing also provide a natural explanation why these $D$ wave vector mesons are difficult to find at $e^+e^-$ collider.
From Tab.~\ref{mixTab}, one can also read off the open channel strong decay widths after considering $S-D$ mixing for states above threshold.
However, one cannot compare the imaginary part of $M_\text{comp}$ directly with experimental data
because its real part (which is the Breit-Wigner mass) does not equal the experimental mass,
thus their phase space for $B\bar{B}$ are different from experiment.

\begin{figure}[H]
  \centering
  \includegraphics[width=0.65\textwidth]{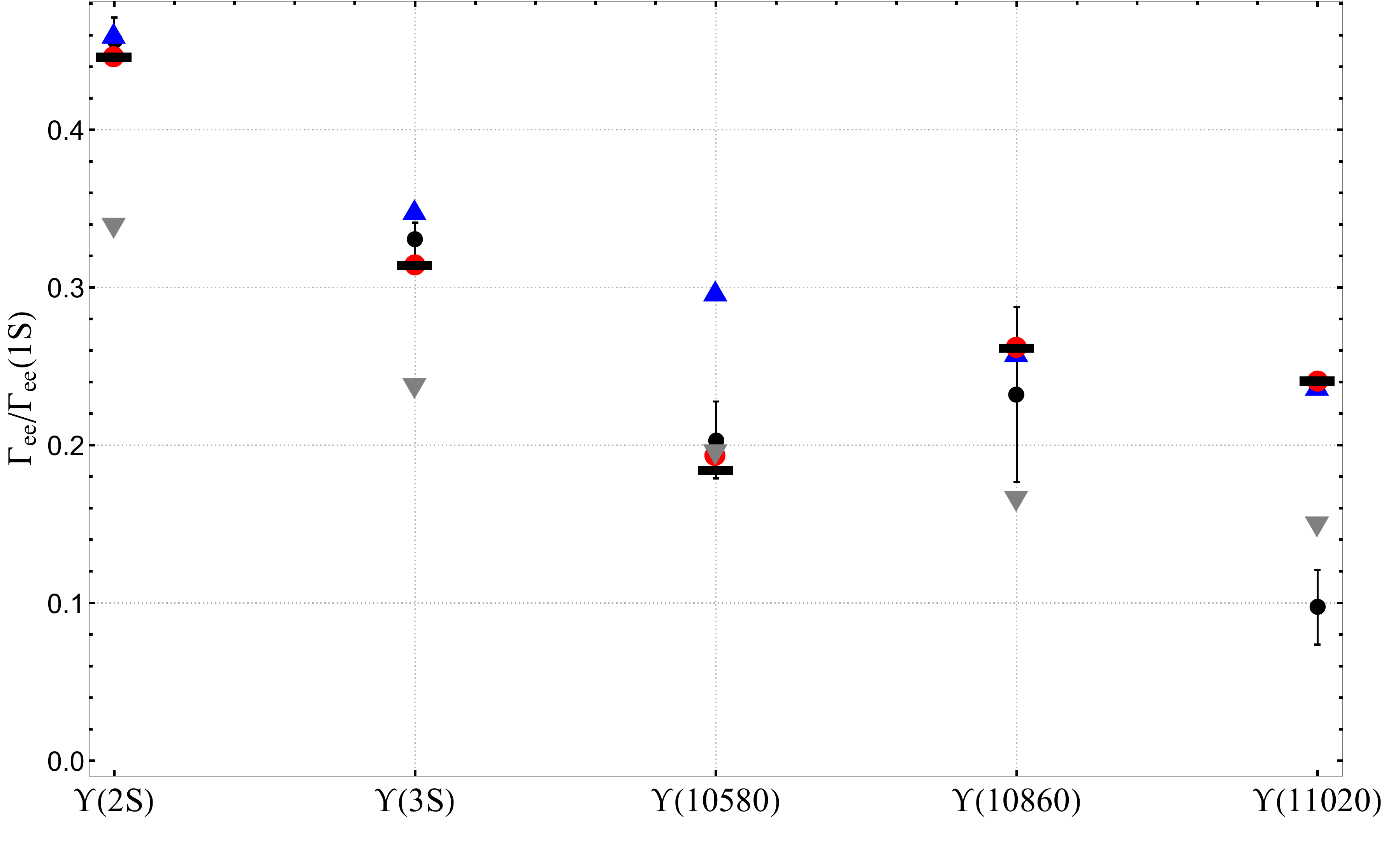}\\
  \caption{Comparison of $\Gamma_{ee}/\Gamma_{ee}(1S)$ between different models.
  Results of the Cornell potential with our parameters and parameters of Ref.~\cite{Liu:2011yp}
  are respectively represented by blue regular and gray inverted triangles.
  Red dots and black rectangles respectively denote the predictions of Eq.~\ref{SDMixing} with and without neglecting the imaginary part.
  Black dots with error bars are values taken from Particle Data Group (PDG)~\cite{Agashe:2014kda}.
  }
  \label{eeDecayPlot}
\end{figure}

A natural and direct consequence of non-negligible $S-D$ mixing
is the suppression of $\Gamma_{ee}(S)$ or enhancement of $\Gamma_{ee}(D)$.
As can be seen from Fig.~\ref{eeDecayPlot},
the dielectric decay width of $\Upsilon(10580)$ and $\Upsilon(11020)$ is highly suppressed experimentally.
Under the assumption that $\Upsilon(10580)$ and $\Upsilon(11020)$ are $S$ wave dominate states,
one may be tempted to introduce a large $S-D$ mixing angle for these highly excited states (see, e.g., Ref.~\cite{Badalian:2009bu}).

The unexpected large central value of $\Gamma_{ee}(\Upsilon(10860))$ seems to favor a small $S-D$ mixing angle,
however, due to its large errors,
a mixing angle as large as $27^\circ$ can also reproduce the data which gives the decay width lies at the lower bound~\cite{Badalian:2009bu}.
Of course, more precise measurement of $\Upsilon(10860)$'s dielectric decay will tell us
whether the claim of large $S-D$ mixing is correct or not,
if $S-D$ mixing is fully responsible for this suppression.

As can be seen from Tab.~\ref{mixTab},
except for the $4S-3D$ case, which will be discussed shortly,
we get a rather small mixing angle not only for below threshold states,
but also for highly excited states.
It seems that the coupled channel formalism cannot explain the suppression of $\Gamma_{ee}$.
However, we want to point it out that $S-D$ mixing is not the only way to suppress $\Gamma_{ee}(S)$.
Even though the $S-D$ mixing angles are small,
$\Gamma_{ee}$ can still be suppressed by the $B$ meson continuum.

The suppression due to $B$ meson continuum is not difficult to understand.
If the conventional mesons have non-negligible components of $B$ meson pairs,
these meson pairs have to undergo one more $\3P0$ vertex before annihilating into $e^+e^-$ pairs.
Since the Hamiltonian is small compared with Cornell potential,
it is reasonable to discard the contribution of these meson pairs, then $\Gamma_{ee}$ is suppressed.
This suppression is universal both for $S$ and $D$ wave vector mesons,
in contrast to $S-D$ mixing, which enhances the $\Gamma_{ee}(D)$.

We take into account both $S-D$ mixing (see Tab.~\ref{mixTab}) and $B$ meson pair suppression mechanism (see Tab.~\ref{ratioTab}) in this work,
and the results of $\Gamma_{ee}/\Gamma_{ee}(1S)$ are shown in Fig.~\ref{eeDecayPlot}.

From Fig.~\ref{eeDecayPlot}, one can see that the coupled channel results agree well with experiment except for $\Upsilon(11020)$.
The large suppression of $\Gamma_{ee}(\Upsilon(10580))$ deserves more explanation.
From quark model's perspective, $\Gamma_{ee}(\Upsilon(10580))$ is suggested to be a $4S$ or $3D$ state.
If it is $S$ wave dominate, the mixing angle is about $9^\circ$,
which is still not big enough to reproduce $\Gamma_{ee}$.
In fact, according to our calculation, the major suppression comes from the $B$ meson pairs.

Even though $\Upsilon(10580)$ is above $B\bar{B}$ threshold,
it is still below and quite close to $B\bar{B}^*$ threshold.
As $h_b(3P),\chi_{b0}(3P)$, and $\chi_{b1}(3P)$ are shown to us in Tab.~\ref{ratioTab},
closer to the threshold means bigger probabilities of $B$ meson continuum.
Like the authors did in Ref.~\cite{Heikkila:1983wd},
we neglect $B\bar{B}$'s probability and work out the probabilities of other channels.
The probabilities of $B\bar{B}^*+h.c., B^*\bar{B}^*, B_s\bar{B}_s, B_s\bar{B}_s^*+h.c., B_s^*\bar{B}^*_s$ are
$20.06\%, 11.7\%, 0.125\%, 0.37\%, 0.51\%$, respectively, that means $P_{b\bar{b}}=67.2\%$.
So as an estimation, one will get only two thirds of the decay width predicted by the potential model.

For $3D$ dominant states with Breit-Wigner mass $10.731$ GeV in Tab.~\ref{mixTab},
its large mixing angle may grasp one's attention.
Because that it is more difficult to generate at $e^+e^-$ collider compared with $4S$ dominant states,
and its mass is $80$ MeV heavier than $4S$ dominant states,
we do not consider it as $\Upsilon(10580)$.
In the $4S-3D$ mixing case,
because of the oscillation behavior of $\Delta M_f$ and $\bra{\psi_f}H_I\ket{\psi_i}$,
there is one more pair of solutions of $M$ in Eq.~(\ref{SDMixing}) with GEM.
For $4S$ dominate state, $m_{\text{comp}}=10.673 +0.0989 i, m_{\text{real}}=10.675, \theta=18.18^\circ$,
and for $3D$ dominate state, $m_{\text{comp}}=10.718 +0.0441 i, m_{\text{real}}=10.7233, \theta=18.4^\circ$.
With the same reasons we also do not consider it as $\Upsilon(10580)$.

Another interesting detail of the $B$ meson continuum is
the slightly increased ratio of $\Gamma_{ee}/\Gamma_{ee}(1S)$.
The mixing angles of $5S-4D$ and $6S-5D$ are so small that $\Gamma_{ee}$ barely change,
nevertheless, due to the small $B$ meson continuum component of $\Upsilon(1S)$,
$\Gamma_{ee}(1S)$ will be suppressed about 0.013, as a consequence,
the ratio $\Gamma_{ee}/\Gamma_{ee}(1S)$ becomes slightly larger after considering coupled-channel effects.

For $\Upsilon (10860)$ and $\Upsilon(11020)$,
all the ground state $B$ meson channels are open.
We can no longer deduce the probabilities of these $B$ meson pairs with Eq.~(\ref{SDMixing}).
There is no $B$ meson continuum suppression in this work.
This can reproduce the dielectric decay width of $\Upsilon (10860)$ but not $\Upsilon(11020)$.

As shown in Fig.~\ref{eeDecayPlot},
there is a notable discrepancy between our calculation and experiment on $\Upsilon(11020)$ dielectric decay width.
This issue may come from the two assumptions we use to simplify the calculation.
One is that we only consider the mixing between $6S$ and $5D$.
In fact, with the increase of radial quantum number,
the energy levels of $S$ or $D$ wave will become denser,
so the mixing may exist between several $S$ and $D$ wave states.
Another is the probabilities of excited $B$ meson pairs are neglected.
This may also cause problems.
For example, the $\Upsilon(11020)$ is only $26$ MeV lighter than $m_{B^*}+m_{B_1}$,
so a large suppression of $P_{b\bar{b}}$ is naturally expected,
causing a large suppression of the dielectric decay width.

It is possible to distinguish $S-D$ mixing and $B$ meson pairs suppression mechanism by the measurement of the radiative decay.
Theoretically, the $E1$ transition can be represented by~\cite{Kwong:1988ae,Rosner:2001nm,Li:2009zu}
\be
\Gamma(n^{2S+1}L_J\to n'^{2S'+1}L'_{J'}+\gamma)=\frac{4}{3}C_{fi}\delta_{SS'}e^2_{b}\alpha |\bra{f}r\ket{i}|^2E^3_{\gamma},
\label{e1Decay}
\ee
where $e_b=-\frac{1}{3}$. $\alpha$ and $ E_\gamma$ respectively denote the fine structure constant and the energy of the emitted photon.
$\bra{f}r\ket{i}$ and $C_{fi}$ are represented by
\ba
\bra{f}r\ket{i}&=&\int_0^\infty R_f(r)R_i(r) r^3 dr, \label{rOverlap} \\
C_{fi}&=&\max(L,L')(2J'+1)
\left\{
\begin{array}{ccc}
L'& J'& S\\
J & L &1
\end{array}
\right\}^2.
\ea

From Eq.~(\ref{e1Decay}), we have
\ba
r_\gamma(S)&:=&\frac{\Gamma(\Upsilon(S)\to\chi_{b2}(1P)+\gamma)}{\Gamma(\Upsilon(S)\to\chi_{b0}(1P)+\gamma)}=5\left(\frac{E_{\gamma2}}{E_{\gamma0}}\right)^3, \label{rGammaS}\\
r_\gamma(D)&:=&\frac{\Gamma(\Upsilon_1(D)\to\chi_{b2}(1P)+\gamma)}{\Gamma(\Upsilon_1(D)\to\chi_{b0}(1P)+\gamma)}=\frac{1}{20}\left(\frac{E_{\gamma2}}{E_{\gamma0}}\right)^3,\label{rGammaD}
\ea
where $E_{\gamma2}$ and $E_{\gamma0}$ respectively represent the photon energy of
$V\to\chi_{b2}+\gamma$ and $V\to\chi_{b0}+\gamma$. ($V$ stands for initial vector state.)

From PDG data~\cite{Agashe:2014kda}, we have $r_\gamma(2S)=1.91\pm0.29$ and $r_\gamma(3S)=3.82\pm1.05$,
and the theoretical predictions of  Eq.~(\ref{rGammaS}) and Eq.~(\ref{rGammaD}) are $r_\gamma(2S)=1.57$, $r_\gamma(3S)=3.6$,
$r_\gamma(1D)=0.0157$ and $r_\gamma(2D)=0.036$.
So it is reasonable to treat $\Upsilon(2S)$ and $\Upsilon(3S)$ as pure $S$ wave,
and our conclusion of small $S-D$ mixing angle is consistent with experiment for vector bottomonia below threshold.

If the all vector bottomonia observed are $S$ wave dominant,
the small $\Gamma_{ee}$ of $\Upsilon(10580)$ and $\Upsilon(11020)$ naturally requires a large mixing angle
under $S-D$ mixing mechanism, causing a large suppression of $r_\gamma(S)$.
On the contrary, $B$ meson continuum suppresses $P_{b\bar{b}}$, leaving the ratio $r_\gamma$ unchanged.
Given that the deduced $S-D$ mixing angle is small, we expect a large $r_\gamma$.
Unfortunately, the data on the radiative decay widths of
$\Upsilon(10580), \Upsilon(10860)$, and $\Upsilon(11020)$ is still not available so far.
A precise measurement of radiative decay will definitely tell us more about their internal structures.

For states above threshold,
the predicted spectra and decay widths agree not very well with experimental data.
There are two reasons to cause this issue.
As with most work, the meson loops of excited $B$ mesons are ignored,
however, this assumption may be not appropriate for highly excited states.
For example, $\Upsilon(11020)$ is already $20$ MeV heavier than $B_1 \bar{B}$ threshold.
The second reason comes from the nonrelativistic approximation of our bare mass.
In principle, relativistic corrections will be more important when the binding energy goes high,
so both the wave functions and the bare masses will change accordingly.
However, including contributions of excited $B$ mesons
and refitting the spectrum and decay widths involves much more work, which lies beyond this work.
It still remains a challenge to reproduce the spectra and dielectric or hadronic decay patterns.

\section{Summary}\label{summary}
In this paper, we make a thorough and precise calculation of coupled-channel effects in the framework of $\3P0$ model with GEM
for the bottomonium.
The results of the spectrum, open channel strong decays, probabilities of the $B$ meson continuum,
the $S-D$ mixing and the vector meson's dielectric decays are explicitly shown.
In order to study the near threshold effects,
we also plot the mass dependence of the mass shift and open channel decay widths for pure $S$ and $D$ wave vector mesons.

For $\Upsilon(4S)$, the decay width of GEM can be two times as large as SHO over a wide energy range,
and the mass shift is around three times as large.
These big deviations indicate that SHO is not a good approximation for near threshold states,
even though the oscillator parameters are carefully selected to reproduce the root mean square radius of the corresponding mesons.

With the consideration of coupled-channel effects, we get small $S-D$ mixing angles except for $\Upsilon(4S)$.
We point it out that, for $S$ wave dominant vector states,
$S-D$ mixing is not the only mechanism to suppress their dielectric decay widths,
$B$ meson continuum can also lead to the suppression.
With this $B\bar{B}$ suppression mechanism at hand,
we still succeed to reproduce the dielectric decays of vector bottomonia except for $\Upsilon(11020)$.
The deviation of the spectrum and decays between our predictions and experimental measurements
may be due to the neglect of excited $B$ meson continuum in coupled-channel effects or
the nonrelativistic approximation in the quenched limit.

$S-D$ mixing will cause the suppression of the ratio in Eq.~(\ref{rGammaS}) for $S$ wave dominant state,
on the contrary, the $B$ meson continuum does not change this ratio.
We suggest \emph{BABAR} and Belle to make precise measurements on the radiative decays of
 $\Upsilon(10580),\Upsilon(10860)$, and $\Upsilon(11020)$ to distinguish these two effects.

\section*{Acknowledgements}
The authors are grateful to
Meng~Ce, Gui-Jun~Ding, David~R.~Entem, Feng-Kun~Guo, Yu.~S.~Kalashnikova, Bilal~Masud, Jia-Lun~Ping and E.~Santopinto
for useful discussions and suggestions.
This work is supported by the National Natural Science Foundation of China under Grants No.~11261130311 (CRC110 by DFG and NSFC).
M. Naeem~Anwar is supported by CAS-TWAS President's Fellowship for International Ph.D Students.

\providecommand{\href}[2]{#2}\begingroup\raggedright\endgroup

\end{document}